\documentclass[a4]{article}

\usepackage{latexsym}
\usepackage{amssymb}
\usepackage{amsthm}
\usepackage{amsmath}
\usepackage{amsfonts}

\usepackage[dvips]{graphicx}

\newcommand{\sikib}{\begin{eqnarray}}
\newcommand{\sikie}{\end{eqnarray}}
\newcommand{\sikibnon}{\begin{eqnarray*}}
\newcommand{\sikienon}{\end{eqnarray*}}
\newcommand{\cenb}{\begin{center}}
\newcommand{\cene}{\end{center}}
\newcommand{\bsplit}{\begin{equation} \begin{split}}
\newcommand{\esplit}{\end{split} \end{equation}}

\newcommand{\js}{\Sigma}
\newcommand{\sg}{\kappa}

\newcommand{\rt}{r^{\star}}
\newcommand{\Rt}{R^{\star}}

\newcommand{\eb}{\bar{\eta}}

\newcommand{\SFD}{\Phi}
\newcommand{\sfd}{\phi}
\newcommand{\sfdWX}{\tilde{\phi}}

\newcommand{\psiWX}{\tilde{\psi}}
\newcommand{\PsiWX}{\tilde{\Psi}}
\newcommand{\PsiWXbyTR}{\hat{\Psi}}

\newcommand{\vac}{\left|0\right>}
\newcommand{\dvac}{\left<0\right|}
\newcommand{\sgn}{\text{sgn}}
\newcommand{\pv}{\text{pv}}

\newcommand{\E}{\text{Erfc}}
\newcommand{\DE}{\Delta\text{E}}
\newcommand{\Han}{\text{H}}

\newcommand{\od}[2]{ \frac{d #1}{d #2} }
\newcommand{\inpro}[2]{ \left( #1 \, , \, #2 \right) }

\title{Hawking radiation in the swiss cheese universe}

\author{ Hiromi SAIDA
\footnote{E-mail: saida@scisv.sci.osaka-cu.ac.jp /
                  saida@phys.h.kyoto-u.ac.jp} \,
\footnote{From Apr.2002 attached to Department of Mathematics and 
Physics, Graduate School of Science, Osaka City University. 
Sugimoto 3-3-138, Sumiyoshi-ku, Osaka 558-8585, Japan.}\\[3mm]
  Department of Natural Environment, \\
  Graduate School of Human and Environmental Studies, 
  Kyoto University, \\ 
  Kyoto 606-8501, Japan}

\date{}

\begin{document}

\maketitle


\begin{abstract}

The Hawking radiation forms the essential basis of the black hole 
thermodynamics. The black hole thermodynamics denotes a nice 
correspondence between black hole kinematics and the laws of ordinary 
thermodynamics, but has been so far considered only in an 
asymptotically flat case. {\it Does such the correspondence rely 
strongly on the feature of the gravity vanishing at the infinity?} In 
order to resolve this question, it should be considered for the first 
to extend the Hawking radiation to a case with a dynamical boundary 
condition like an expanding universe. Therefore the Hawking radiation 
in an expanding universe is discussed in this paper. As a concrete 
model of a black hole in an expanding universe, we use the swiss 
cheese universe which is the spacetime including a Schwarzschild black 
hole in the Friedmann-Robertson-Walker universe. Further for 
simplicity, our calculation is performed in two dimension. The 
resultant spectrum of the Hawking radiation measured by a comoving 
observer is generally different from a thermal one. We find that the 
qualitative behavior of the non-thermal spectrum is of dumping 
oscillation as a function of the frequency measured by the observer, 
and that the intensity of the Hawking radiation is enhanced by the 
presence of a cosmological expansion. It is appropriate to say that 
{\it a black hole with an asymptotically flat boundary condition 
stays in a lowest energy thermal equilibrium state, and that, once a 
black hole is put into an expanding universe, it is excited to a 
non-equilibrium state and emits its mass energy with stronger 
intensity than a thermal one.}

\end{abstract}

Submitted to the Classical and Quantum Gravity



\section{Introduction}\label{sec-intro}

So far various properties of black hole spacetimes have been 
revealed. Among so many well-known properties of the black hole, one 
of the most remarkable ones is the black hole thermodynamics. This 
tells us a nice correspondence between the classical properties of a 
black hole spacetime and the laws of ordinary thermodynamics 
\cite{ref-thermo}. The black hole thermodynamics is one of the most 
important stages for understanding the nature of a strong gravity. 
However this nice correspondence has been considered only in the 
asymptotically flat case. {\it Does such the correspondence rely 
strongly on the feature of the gravity vanishing at the infinity?} It 
has not been clarified if such the correspondence is true of the case 
with a dynamical boundary condition.
\footnote{It is not too much to say that only a few studies have been 
done on the issue of a black hole in some dynamical background 
\cite{ref-dyn}.}
My general purpose is to {\it extend the black hole thermodynamics to 
a dynamical situation, and hopefully find a non-gravitational system 
which corresponds to a black hole with a dynamical boundary condition 
like an expanding universe.} By the way, the Hawking radiation plays 
the essential role in the black hole thermodynamics to determine 
precisely the temperature of a black hole with asymptotically 
flatness \cite{ref-hr}. Hence, as the first step to attack my 
purpose, this paper is especially designed for the Hawking radiation 
in an expanding universe. 

There is a serious problem of defining a black hole. A black hole is 
defined as the spacetime region excluded from the causal past of the 
future null infinity \cite{ref-he}. The future null infinity for the 
asymptotically flat case is well known, therefore it is rather clear 
if an asymptotically flat spacetime includes a black hole. On the 
other hand, however, under a dynamical boundary condition, it is 
generally quite difficult to extract the asymptotic structure of the 
spacetime directly from the Einstein equation. The black hole cannot 
be defined unless the asymptotic structure is known. In order to 
avoid the problem of defining the black hole in an expanding 
universe, we make use of the {\it swiss cheese universe}. 

The swiss cheese universe (SCU) is the spacetime including a 
Schwarzschild black hole in a Friedmann-Robertson-Walker (FRW) 
universe. It is not the solution obtained by solving directly the 
Einstein equation, but the solution constructed by connecting the 
Schwarzschild spacetime with the FRW universe at a given spherically 
symmetric timelike hypersurface $\js$ by the Israel junction 
condition with requiring no surface energy density confined on $\js$ 
\cite{ref-scu} \cite{ref-israel}. The spatial section of $\js$ which 
is a two dimensional sphere is expanding as seen from the black hole 
side with respect to the Schwarzschild coordinate. So once the two 
sphere on $\js$ was placed outside the gravitational radius 
$R_g = 2GM$ at a given time, the event horizon should last to exist 
after that time. That is, the existence of a black hole is guaranteed 
in the SCU. 

Another technical problem has still been left. Even if with the 
asymptotic flatness, the curvature scattering makes it very difficult 
to obtain a complete form of the Hawking spectrum in four dimensional 
case. Indeed, the curvature scattering is ignored in the statement 
that an asymptotically flat black hole is in a thermal equilibrium 
state. So we also want to ignore the curvature scattering for 
calculating the Hawking radiation in the SCU. By the way, the 
curvature scattering for a minimal coupling massless scalar field 
vanishes on any two dimensional spacetimes due to the conformal 
flatness of the spacetime. Therefore in this paper, we treat the two 
dimensional SCU for simplicity, and introduce the minimal coupling 
massless scalar field for calculating the Hawking radiation. 

In section \ref{sec-scu}, we briefly explain the two dimensional 
``eternal'' swiss cheese universe and show the strategy to obtain the 
Hawking radiation on two dimensional SCU. The behavior of a minimal 
coupling massless scalar field on the eternal SCU is analyzed in 
section \ref{sec-sf}. Section \ref{sec-hr} is devoted to computing 
the Hawking radiation. Summary and discussions are given in section 
\ref{sec-sd}. 

The unit throughout this paper is $c=\hbar=k_B=1$. The metric 
signature is $(-,+,+,+)$ for four dimension and similarly for two 
dimension.


\section{``Eternal'' swiss cheese universe and the strategy}
\label{sec-scu}

\subsection{Two dimensional ``eternal'' SCU}
\label{sec.scu-scu}

The two dimensional SCU should be constructed by connecting the 
Schwarzschild spacetime with the FRW universe at a given timelike 
surface (curve) $\js$. The way of junction is very similar to the 
four dimensional case explained in appendix \ref{app-scu}. Hereafter 
as the terminology, let the word ``BH-side'' denote the spacetime 
region inside $\js$ where the metric is given by 
(\ref{eq-sec.scu.scu-BH.TR}) or (\ref{eq-sec.scu.scu-BH.WX}), and 
``FRW-side'' for the region outside $\js$ where the metric is 
(\ref{eq-sec.scu.scu-FRW.com}) or (\ref{eq-sec.scu.scu-FRW.Cfl}). If 
a quantity $Q$ is measured in the BH-side, we express it as $Q_{BH}$ 
and similarly $Q_F$ in the FRW-side. 

The metric of the BH-side is given by 
 \sikib
  ds_{BH}^2 &=&
    - \left( 1-\frac{R_g}{R} \right) dT^2 
    + \left( 1-\frac{R_g}{R} \right)^{-1} dR^2
    \quad , \text{in TR-system} 
  \label{eq-sec.scu.scu-BH.TR} \\
     &=&
  \frac{ 4R_g^{\,\,3} }{R} \, e^{-R/R_g} \left[ -dW^2 + dX^2 \right] 
    \quad , \text{in WX-system} \, ,
  \label{eq-sec.scu.scu-BH.WX}
 \sikie
where $R_g=2GM$, $G$ is the gravitational constant, and the meaning 
of $M$ is given at the end of this paragraph. The terms ``TR-system'' 
and ``WX-system'' mean respectively the Schwarzschild coordinate and 
the Kruskal-Szekeres coordinate in the BH-side. Because the maximally 
extended Schwarzschild spacetime has two exterior regions of a black 
hole, there can also be two exterior regions in the BH-side of the 
SCU. We distinguish them by calling ``L-region'' and ``R-region''. 
The coordinate transformation between the TR-system and the WX-system 
is given by
 \sikib
  W-X = \mp \exp[-\sg(T-\Rt)] \quad , \quad
  W+X = \pm \exp[ \sg(T+\Rt)] \, ,
 \label{eq-sec.scu.scu-coordtrans.BH}
 \sikie
where the upper signature is for the R-region and the lower for the 
L-region, $\Rt=R+R_g\ln(R/R_g-1)$ and $\sg=1/2R_g$. We set the 
Killing vector $\tilde{\xi}_{BH}$ in the WX-system as
 \sikib
  \tilde{\xi}_{BH} = \partial_W \quad
  \text{, all over the BH-side} \, ,
 \label{eq-sec.scu.scu-kv.WX}
 \sikie
where a tilde is added to the quantity measured in the WX-system. 
Adopting the direction of this Killing vector as the standard of the 
future direction, the Killing vector $\xi_{BH}$ in the TR-system is 
given by
 \sikib
  \xi_{BH} =
   \begin{cases}
     \partial_T & \text{, in the R-region} \\
    -\partial_T & \text{, in the L-region} \, .
   \end{cases}
 \label{eq-sec.scu.scu-kv.TR}
 \sikie
The normalization of the vectors (\ref{eq-sec.scu.scu-kv.WX}) and 
(\ref{eq-sec.scu.scu-kv.TR}) is determined at the asymptotically flat 
region of the full Schwarzschild spacetime which is not present in 
the SCU. However once we accept them as the timelike Killing vector, 
$\sg$ can be interpreted as the surface gravity of the event horizon, 
and further $M$ becomes exactly equal to the Komar mass 
\cite{ref-townsend} evaluated on any spatial section of $\js$. 
Such a Komar mass \cite{ref-townsend} defined using our Killing 
vector (\ref{eq-sec.scu.scu-kv.WX}) or (\ref{eq-sec.scu.scu-kv.TR}) 
is invariant under the deformation of the spatial section of $\js$. 
See the end of appendix \ref{app-scu} for more detailed explanation 
of $\sg$ and $M$.

The metric of the FRW-side in the comoving coordinate is
 \sikib
  ds_F^2 = - dt^2 + a(t)^2 \frac{dr^2}{1-k r^2} \, ,
 \label{eq-sec.scu.scu-FRW.com}
 \sikie
where $k=\pm1$, $0$ is the spatial curvature, and $a(t)$ is the scale 
factor. The comoving spatial coordinate in two dimensional case 
can range over infinity $-\infty<r<\infty$ for open and flat cases 
$k=-1,0$. There is coordinate singularities at $r=\pm1$ for closed 
case $k=1$, but it can be eliminated by the transformation from the 
comoving coordinate $(t,r)$ to the conformal coordinate $(\eta,\rt)$ 
which is given by $d\eta = dt/a(t)$ and $r=\sin\rt$, $\rt$, or 
$\sinh\rt$ for $k=1$, $0$ or $-1$ respectively. Then the metric 
becomes 
\sikib
 ds_F^2 = a(\eta)^2 \left( - d\eta^2 + d\rt\,^2 \right) \, ,
 \label{eq-sec.scu.scu-FRW.Cfl}
\sikie
where $-\infty<\rt<\infty$. Hereafter as the terminology we call the 
conformal coordinate the Cfl-system. The comoving coordinate is 
equivalent to the Cfl-system in the sense that there exists one 
comoving coordinate point $(t,r)$ for each point of the Cfl-system 
$(\eta,\rt)$, where this correspondence is the one-to-one mapping for 
open and flat cases and the onto-mapping for closed case. 

The two dimensional FRW metric does not have any 
timelike Killing vector, but has a conformal timelike Killing vector 
$\xi_F$ which satisfies 
$\xi_{F\, \mu ; \nu} + \xi_{F\, \mu ; \nu} \propto g_{\mu \nu}$, 
 \sikib
  \xi_F =
   \begin{cases}
     \partial_{\eta} & \text{, in the R-region} \\
    -\partial_{\eta} & \text{, in the L-region} \, .
   \end{cases}
 \label{eq-sec.scu.scu-kv.Cfl}
 \sikie

\begin{figure}[t]
  \cenb
   \includegraphics[height=35mm]{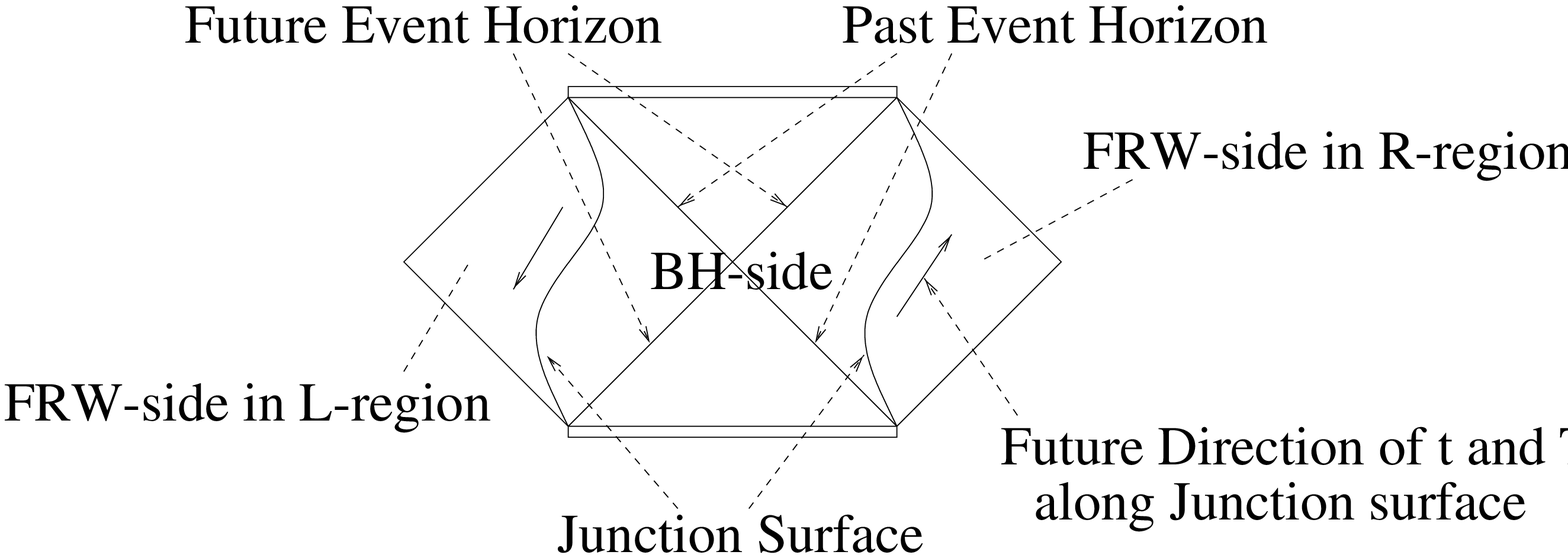}
  \cene
 \caption{{\small Two dimensional eternal swiss cheese universe. The 
junction surface is set outside the gravitational radius so that a 
Black hole exists eternally in the BH-side.}}
 \label{fig-scu.eternal}
 \end{figure}

In order to treat the two dimensional SCU as a simplification of four 
dimensional case, we require the same kinematics of the junction 
surface and the same relation between the temporal coordinates $t$ and 
$T$ as four dimensional case shown in the appendix \ref{app-scu}. 
That is, the location of $\js$ is given by
 \sikib
   \js :
    \begin{cases}
     R(t)=a(t) \, r_0 &, \text{in the BH-side,} \\
     r = r_0 (=\text{const.}) &, \text{in the FRW-side,} 
    \end{cases}
 \label{eq-sec.scu.scu-js}
 \sikie
and the temporal coordinates are related by
 \sikib
   \od{T(t)}{t} = \frac{\sqrt{1-k\,r_0^{\,\, 2}} }{1-R_g/R(t)} \, .
 \label{eq-sec.scu.scu-time}
 \sikie
The unit normal vector to $\js$ is given by the same form as the four 
dimensional case (\ref{eq-app.scu-normal.FRW}) and 
(\ref{eq-app.scu-normal.BH})
 \sikib
   n_F^{\mu} &=&
     \left( 0 \,\, , \,\, \frac{1}{a(t)} \right)
     \quad \text{in Cfl-system} ,
 \label{eq-sec.scu.scu-normal.FRW} \\
   n_{BH}^{\mu} &=&
     \left( \frac{\dot{R}(t)}{1-R_g/R(t)} \,\, , \,\,
            \left(1-\frac{R_g}{R(t)}\right) \dot{T}(t) \right)
     \quad \text{in TR-system} ,
 \label{eq-sec.scu.scu-normal.BH}
 \sikie
where $\dot{Q}=dQ/dt$. Note that the equation 
(\ref{eq-sec.scu.scu-time}) and the normalization of 
(\ref{eq-sec.scu.scu-normal.BH}), ${\bf n}_{BH}^{\,\,\,2}=1$, gives 
the Friedmann equation of dust matter (\ref{eq-app.scu-sfac}). 
However in order to extract the essence of the Hawking radiation in 
an expanding universe in a simple way, we dear to assume that the 
scale factor satisfies $a(t)\,r_0 > R_g$ for $-\infty<t<\infty$. That 
is, we consider the two dimensional SCU in which a black hole can 
exist eternally. The conformal diagram of such the ``eternal'' SCU is 
shown at the figure \ref{fig-scu.eternal}. With this assumption, a 
black hole can be defined without respect to the spatial curvature of 
the FRW-side, and the event horizon bifurcates. Hereafter throughout 
this paper, we adopt the eternal SCU as the background spacetime. The 
validity of the eternal SCU will be discussed at the end of section 
\ref{sec-sd}. 

\subsection{Strategy for the Hawking radiation}
\label{sec.scu-str}

\subsubsection{Choice of the Vacuum state}
\label{sec.scu.str-vac}

As discussed in section \ref{sec-intro}, we introduce a minimal 
coupling massless scalar field $\SFD$ and neglect any back reaction 
of $\SFD$ to the background eternal SCU. The purpose of this paper is 
to compute the Hawking radiation received by a comoving observer in 
the FRW-side. Then, because of the equivalence between the comoving 
coordinate and the Cfl-system, the expectation value of the number 
operator of the quantized scalar field in the Cfl-system attracts our 
interest. 

Here we have to specify on which vacuum state the expectation value 
should be calculated. For the asymptotically flat black hole 
spacetime, there are two candidates for such the vacuum state: the 
Hartle-Hawking state \cite{ref-hh} and the Unruh state 
\cite{ref-unruh}. The former describes that a black hole is 
surrounded by a thermal cavity of temperature $\kappa/2\pi$ and the 
whole system is in equilibrium. On the other hand, the latter 
corresponds to the situation that the environment surrounding the 
black hole is not a thermal cavity but an empty space, that is, the 
black hole is evaporating due to the Hawking radiation without 
absorbing any energy from the empty environment. By the way, it is 
not {\it a priori} known whether the black hole in the SCU emits a 
thermal radiation or not. Therefore it seems appropriate for the SCU 
not to put the black hole in a thermal cavity, but to put in an empty 
environment. Then we search what kind of spectrum is radiated out of 
the black hole in the SCU. From above considerations, we adopt the 
Unruh state as the vacuum state on which the Hawking radiation is 
computed. Further due to the presence of the junction surface, we use 
the following strategy of computing the Hawking radiation.

\subsubsection{First step: construction of the mode functions}
\label{sec.scu.str-first}

The vacuum state is constructed by quantizing the scalar field 
$\SFD$. In order to quantize $\SFD$, we have to obtain the positive 
frequency mode functions. On the SCU, there are two categories 
of mode functions: one of them is a collection of the mode solutions 
which is monochromatic in the BH-side but a superposition of some 
monochromatic modes in the FRW-side because of the junction 
conditions of $\SFD$, (\ref{eq-app.jc-jcsf.sf}) and 
(\ref{eq-app.jc-jcsf.diff}) given in appendix \ref{app-jc}. Another 
category is of the monochromatic mode in the FRW-side but a 
superposition in the BH-side. And note that a mode function generally 
takes a different functional form according to the coordinate system 
in which the Klein-Gordon equation is solved. Then we search the 
following three modes;
 \begin{enumerate}
  \item Monochromatic mode in BH-side with respect to the WX-system, 
which we call the WX-mode
  \item Monochromatic mode in BH-side with respect to the TR-system, 
which we call the TR-mode
  \item Monochromatic mode in FRW-side with respect to the 
Cfl-system, which we call the Cfl-mode
 \end{enumerate}
Because our background is two dimensional and $\SFD$ is massless and 
has no coupling to gravity, the curvature scattering of $\SFD$ never 
occur. Therefore what we have to take care is the possibility of the 
reflection of $\SFD$ at $\js$. Further as seen in the next section, 
it will be revealed that no reflection at $\js$ arises due to the 
special property of our junction conditions, and we can accomplish 
completely the construction of these modes. Then we define the 
positive frequency WX-mode, TR-mode and Cfl-mode, with respect to the 
(conformal) timelike Killing vectors (\ref{eq-sec.scu.scu-kv.WX}), 
(\ref{eq-sec.scu.scu-kv.TR}) and (\ref{eq-sec.scu.scu-kv.Cfl}). That 
is, the positive frequency mode should satisfy 
${\cal L}_{\xi} \SFD = -i \omega \SFD$, where $\omega (>0)$ is the 
frequency.

\subsubsection{Second step: quantization of the scalar field}
\label{sec.scu.str-second}

We quantize the minimal coupling scalar field $\SFD$ in the 
WX-system, the TR-system and the Cfl-system using the mode functions 
obtained in the first step. Then the vacuum states of each mode 
functions are defined. 

The observer is comoving in the FRW-side, so the vacuum state 
appropriate for describing the empty environment surrounding the 
black hole is that of the Cfl-mode. Therefore the Unruh state for the 
SCU should be composed of two vacuum states: one of them is the 
vacuum of the WX-mode $\vac_W$ on the past event horizon and another 
is that of the Cfl-mode $\vac_{\eta}$ on the past null infinity.
\footnote{For asymptotically flat case, the observer rests on $R=$ 
constant, so the vacuum of the TR-mode is chosen on the past null 
infinity.}

By the way consider if the background is four dimensional, then it is 
very difficult to solve the Klein-Gordon equation on the whole 
spacetime due to the curvature scattering. However near the event 
horizon and the null infinity, the solution can be found. Therefore 
for four dimensional case, we cannot help using the Unruh state as 
our vacuum state on which the Hawking radiation is calculated. 
However on the two dimensional eternal SCU, the exact form of the 
mode function can be constructed on the whole spacetime in the first 
step. So due to the absence of the curvature scattering and the 
reflection at $\js$, the out-going modes radiated from the past event 
horizon are received by the comoving observer without loss of the 
energy of the out-going flux, and the in-going mode from the past 
null infinity never reach the observer. That is, the vacuum 
$\vac_{\eta}$ on the past null infinity has no effect on the Hawking 
radiation received by the comoving observer. Consequently it is 
enough for us to prepare the vacuum state $\vac_W$ at one Cauchy 
surface as the vacuum state on which the Hawking radiation is 
calculated. Hence in the following sections, we do not take care 
about the vacuum $\vac_{\eta}$ in the Unruh state, and aim to 
calculate just the Bogoljubov transformation from the WX-mode to the 
Cfl-mode. This can be understood as the particle of the Cfl-mode 
created in the vacuum state of the WX-mode. 

To proceed this calculation, we make use of another intermediate 
transformation, that is, we perform the successive Bogoljubov 
transformation from the WX-mode to the TR-mode, and from the TR-mode 
to the Cfl-mode. Using this successive transformation, the difference 
between the asymptotically flat case and our SCU case should be 
clarified. The first transformation from the WX-mode to the TR-mode 
gives the thermal Hawking spectrum since the bifurcate event horizon 
of the eternal SCU lets us to make the same discussion given in the 
reference \cite{ref-unruh}. Then the deviation from the 
asymptotically flat case should arise from the second transformation 
from the TR-mode to the Cfl-mode. 

 \begin{figure}[t]
  \cenb
   \includegraphics[height=32mm]{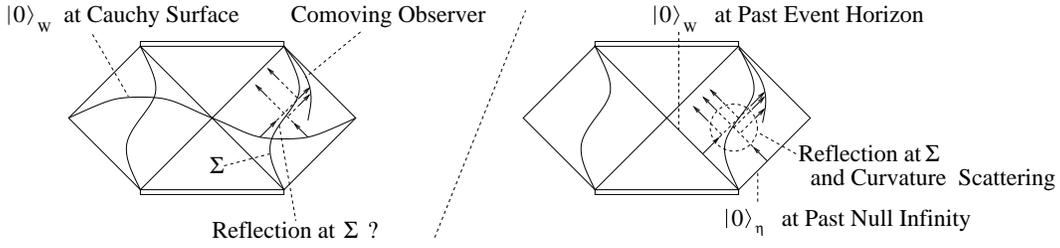}
  \cene
 \caption{{\small Right figure shows the vacuum state for four 
dimensional case. Right one is of two dimensional case. In section 
\ref{sec-sf} it will be found that no reflection takes place for 
two dimensional case, and it is enough to prepare the vacuum $\vac_W$ 
on a Cauchy surface for computing the Hawking radiation in the two 
dimensional SCU.}}
 \label{fig-scu.vac}
 \end{figure}


\section{Massless scalar field in the eternal SCU}
\label{sec-sf}

\subsection{Classical mode function}
\label{sec.sf-mf}

\subsubsection{Preparations}
\label{sec.sf.mf-prep}

We seek the positive frequency TR-mode, WX-mode and Cfl-mode on the 
SCU. However a positive frequency mode is out-going for positive wave 
number, and in-going for negative wave number. Therefore we look for 
the out-going and in-going modes. Those modes can be constructed by 
connecting the mode function obtained on the full Schwarzschild 
spacetime with that obtained on the full FRW spacetime at $\js$. 
Because $\SFD$ is a minimal coupling massless field, it is natural to 
require that no potential of $\SFD$ is confined at $\js$, that is, 
the field equation of the form $\square \SFD=0$ holds even at $\js$. 
This requirement determines the junction conditions of $\SFD$ at 
$\js$, 
 \sikib
  \left. \SFD_{BH} \right|_{\js} = \left. \SFD_{FRW} \right|_{\js}
 \quad , \quad
  \left. \partial_{n} \SFD_{BH} \right|_{\js}
   = \left. \partial_{n} \SFD_{FRW} \right|_{\js} \, ,
 \label{eq-sec.sf.mf.prep-jc}
 \sikie
where $\partial_{n} = n^{\mu}\partial_{\mu}$ and $n^{\mu}$ is the 
unit normal to $\js$, (\ref{eq-sec.scu.scu-normal.FRW}) or 
(\ref{eq-sec.scu.scu-normal.BH}). The detailed derivation of these 
conditions is given in the appendix \ref{app-jc}.

The mode functions satisfying $\square \SFD=0$ on the full 
Schwarzschild spacetime are,
 \sikib
  \sfd_{BH , \Omega}^{(\pm)} &=&
    \frac{1}{\sqrt{4\pi|\Omega|}} \times
   \begin{cases}
     e^{-i \, \Omega \, (T - \Rt)} &, \text{out-going} \\
     e^{-i \, \Omega \, (T + \Rt)} &, \text{in-going}
   \end{cases}
 \quad \text{in TR-system} \, ,
 \label{eq-sec.sf.mf.prep-TRmode} \\
  \sfdWX_{BH , \tilde{\Omega}}^{(\pm)} &=&
    \frac{1}{\sqrt{4\pi|\tilde{\Omega|}}} \times
   \begin{cases}
     e^{-i \, \tilde{\Omega} \, (W-X)} &, \text{out-going} \\
     e^{-i \, \tilde{\Omega} \, (W+X)} &, \text{in-going}
   \end{cases}
 \quad \text{in WX-system} \, ,
 \label{eq-sec.sf.mf.prep-WXmode}
 \sikie
where the upper suffix ``$+$'' denotes the out-going mode and the 
lower ``$-$'' is for the in-going one, and the normalization factor 
is determined by an ordinary inner product of the full Schwarzschild 
spacetime. The mode functions on the full FRW spacetime are 
 \sikib
  \sfd_{F , \omega}^{(1,2)} =
    \frac{1}{\sqrt{4\pi|\omega|}} \times
   \begin{cases}
     e^{-i \, \omega \, (\eta-\rt)} &, \text{out-going} \\
     e^{-i \, \omega \, (\eta+\rt)} &, \text{in-going}
   \end{cases}
 \quad \text{in Cfl-system} \, ,
 \label{eq-sec.sf.mf.prep-Cflmode}
 \sikie
where the upper suffix ``$1$'' denotes the out-going mode and the 
lower ``$2$'' is for the in-going one, and the normalization factor 
is determined by an ordinary inner product. Even if we solve 
$\square \SFD=0$ using the comoving coordinate, the same mode 
function is obtained. For later convenience, we introduce the 
notations:
 \sikib
  f_{F,\omega}(\eta) \equiv e^{-i \omega \eta}
   \quad \text{and} \quad
  h_{F,\omega}^{(1,2)}(\rt) \equiv
    \frac{1}{\sqrt{4\pi|\omega|}} e^{\pm i \omega \rt}
 \label{eq-sec.sf.mf.prep-Cfl.sep} \, .
 \sikie
Note that, because of the assumption of the eternal SCU, the temporal 
part $f_{F,\omega}(\eta)$ can span a complete orthogonal set with 
respect to $\eta \in (-\infty , +\infty)$ and 
$\omega \in (-\infty , +\infty)$ as 
 \sikib
  \int_{-\infty}^{\infty} d\eta \,
         f_{F,\omega}^{\ast}(\eta) \, f_{F,\omega'}(\eta)
    = 2\pi \, \delta(\omega-\omega')
  \quad , \quad
  \int_{-\infty}^{\infty} d\omega \,
         f_{F,\omega}^{\ast}(\eta) \, f_{F,\omega}(\eta')
    = 2\pi \, \delta(\eta-\eta')  \, .
 \label{eq-sec.sf.mf.prep-Cfl.comp} 
 \sikie

Referring to the ordinary definition of the inner products on the 
full Schwarzschild and FRW spacetimes, we define that on the SCU, 
 \sikib
  \inpro{\psi_1}{\psi_2} =
    i \int_{-\infty}^{\Rt(\eta)} d\Rt 
      \psi_1^{\ast}
        \! \stackrel{\leftrightarrow}{\partial_p} \! \psi_2
  + i \int_{\rt_0}^{\infty} d\rt
      \psi_1^{\ast}
        \! \stackrel{\leftrightarrow}{\partial_p} \! \psi_2 \, ,
 \label{eq-sec.sf.mf.prep-inpro}
 \sikie
where $\psi$ denotes the solution of $\square \SFD=0$ on the SCU, 
$\partial_p = (1/|\xi|)\xi^{\mu}\partial_{\mu}$, $\xi$ denotes 
the (conformal) Killing vectors (\ref{eq-sec.scu.scu-kv.Cfl}) and 
(\ref{eq-sec.scu.scu-kv.TR}) or (\ref{eq-sec.scu.scu-kv.WX}), and 
$\rt_0$ is the coordinate value of $\rt$ at the junction surface 
$\js$. This inner product is defined on the spatial surface of 
$\eta=$ constant in the SCU.

\subsubsection{Out-going and in-going TR-modes and WX-modes}
\label{sec.sf.mf-TRWX}

\subsubsection*{TR-mode}

As mentioned at the subsection \ref{sec.scu.str-first}, the out-going 
and in-going TR-modes are monochromatic in the BH-side but a 
superposition of some monochromatic modes in the FRW-side. Then we 
can express the out-going TR-mode $\psi_{BH , \Omega}^{(+)}$ and the 
in-going one $\psi_{BH , \Omega}^{(-)}$ as
 \sikib
  \psi_{BH,\Omega}^{(+)} &=& 
  \begin{cases}
   \sfd_{BH,\Omega}^{(+)}
     &, \text{in BH-side} \\
   \int_{-\infty}^{\infty} d\omega
      \left[ A_{\Omega \omega}^{(1)} \, \sfd_{F , \omega}^{(1)}
           + A_{\Omega \omega}^{(2)} \, \sfd_{F , \omega}^{(2)}
      \right] &, \text{in FRW-side} \, ,
  \end{cases}
  \label{eq-sec.sf.mf.TRWX-outTR} \\
 \psi_{BH,\Omega}^{(-)} &=& 
  \begin{cases}
   \sfd_{BH,\Omega}^{(-)}
     &, \text{in BH-side} \\
   \int_{-\infty}^{\infty} d\omega
      \left[ B_{\Omega \omega}^{(1)} \, \sfd_{F , \omega}^{(1)}
           + B_{\Omega \omega}^{(2)} \, \sfd_{F , \omega}^{(2)}
      \right] &, \text{in FRW-side} \, ,
  \end{cases}
  \label{eq-sec.sf.mf.TRWX-inTR}
 \sikie
where the junction coefficients $A^{(1,2)}$ and $B^{(1,2)}$ are 
determined by the junction conditions (\ref{eq-sec.sf.mf.prep-jc}), 
which become
 \begin{equation}
 \begin{split}
  \sfd_{BH,\Omega}^{(+)}(T(\eta),R(\eta)) \,\, &= \,\,
    \int_{-\infty}^{\infty} d\omega
    \left[
     A_{\Omega \omega}^{(1)} \,
       f_{F,\omega}(\eta) \, h_{F,\omega}^{(1)}(r_0) +
     A_{\Omega \omega}^{(2)} \,
       f_{F,\omega}(\eta) \, h_{F,\omega}^{(2)}(r_0)
    \right]  \, , \\
  \partial_n \sfd_{BH,\Omega}^{(+)}(T(\eta),R(\eta)) \,\, &= \,\,
    \frac{1}{a(\eta)} \int_{-\infty}^{\infty} d\omega
    \left[
     A_{\Omega \omega}^{(1)} \,
       f_{F,\omega}(\eta) \, \od{\,h_{F,\omega}^{(1)}}{\rt}(r_0) +
     A_{\Omega \omega}^{(2)} \,
       f_{F,\omega}(\eta) \, \od{\,h_{F,\omega}^{(2)}}{\rt}(r_0)
    \right] \, , \\
 \end{split}
 \label{eq-sec.sf.mf.TRWX-jcsf}
 \end{equation}
and the similar equations for $B^{(1,2)}$ holds, where 
$(T(\eta), R(\eta))$ is the coordinate point on $\js$ in the 
TR-system, and note that $(dh_F/d\rt)|_{r=r_0} =$ constant. 

Here in this subsection, let us calculate to obtain $A^{(1,2)}$, then 
the similar calculation gives $B^{(1,2)}$. Using the completeness 
(\ref{eq-sec.sf.mf.prep-Cfl.comp}), we obtain the integral 
representations
 \begin{equation}
 \begin{split}
  A_{\Omega \omega}^{(1)} \,\,\, &= \,\,\,
   \sgn(\omega) \,
    \frac{e^{-i \omega \rt_0}}{4\pi \sqrt{|\Omega \omega|}} \,
     \left[
       I_A(\Omega,\omega) - J_A(\Omega,\omega)
     \right] \\
  A_{\Omega \omega}^{(2)} \,\,\, &= \,\,\,
   \sgn(\omega) \,
    \frac{e^{i \omega \rt_0}}{4\pi \sqrt{|\Omega \omega|}} \,
     \left[
       I_A(\Omega,\omega) + J_A(\Omega,\omega)
     \right] \, ,
 \end{split}
 \label{eq-sec.sf.mf.TRWX-coeff.out}
 \end{equation}
where $\sgn(\omega)=\omega/|\omega|$. Here $I_A$ and $J_A$ are given 
by
 \sikib
  I_A(\Omega,\omega) =
   \omega \, \int_{-\infty}^{\infty} \, d\eta \,
             e^{i \omega \eta -i \, \Omega [T(\eta)-\Rt(\eta)]}
  \quad , \quad
  J_A(\Omega,\omega) =
   \Omega \, \int_{-\infty}^{\infty} \, d\eta \, P_A(\eta) \,
             e^{i \omega \eta -i \, \Omega [T(\eta)-\Rt(\eta)]} \, ,
 \label{eq-sec.sf.mf.TRWX-int.out}
 \sikie
where 
 \sikibnon
  P_A(\eta) = \left(
    \frac{a'(\eta) \, r_0}{a(\eta) \sqrt{1-k\,r_0^{\,\,2}} }
    -1 \right) \, \od{T(\eta)}{\eta} \, ,
 \sikienon 
and $a'=da/d\eta $. Once the scale factor is specified, the 
junction coefficients can be determined in principle. One may think 
that the out-going TR-mode is a superposition of the out-going and 
in-going modes in the FRW-side, and so is the in-going TR-mode. 
However it is not the case. To show it, it is essential to notice 
that
 \sikib
  \od{}{\eta}\, e^{i \omega \eta -i \, \Omega [T(\eta)-\Rt(\eta)]}
   &=& i \, \left[ \omega + \Omega P_A(\eta) \right] \,
       e^{i \omega \eta -i \, \Omega [T(\eta)-\Rt(\eta)]} \, ,
 \label{eq-sec.sf.mf.TRWX-phase.diff}
 \sikie
which gives
 \sikib
  I_A + J_A =
    -i \, \left. e^{i \omega \eta -i\, \Omega [T(\eta)-\Rt(\eta)]}
          \right|_{\eta=-\infty}^{\eta=\infty} \, .
 \label{eq-sec.sf.mf.TRWX-int.rel}
 \sikie
Then we obtain the out-going TR-mode in the FRW-side to be
 \sikib
  \psi_{BH,\Omega}^{(+)} \,\, \text{in FRW-side}
 = \int_{-\infty}^{\infty} d\omega \, \left(
       A_{\Omega \omega}^{(1)} \, \sfd_{F,\omega}^{(1)} \right)
 + \sgn(\eb) \, \frac{\pi}{(4\pi)^{3/2}\sqrt{|\Omega|} } \,
       \left. e^{-i \, \Omega [T(\eb) - \Rt(\eb)]}
       \right|_{\eb=-\infty}^{\eb=\infty} \, ,
  \label{eq-sec.sf.mf.TRWX-osc.out}
 \sikie
where $\int_{-\infty}^{\infty}dx \,(1/x) \, \exp(\pm i x)=\pm i \pi$ 
is used. As discussed in the appendix \ref{app-terms}, the second 
term should be zero, and we can find $I_A=-J_A$ due to the equations 
(\ref{eq-sec.sf.mf.TRWX-int.rel}). 

The same way of calculation done so far holds for the other junction 
coefficients $B^{(1,2)}$. Then the out-going and in-going TR-modes 
are obtained to be
 \sikib
  \psi_{BH,\Omega}^{(+)} &=& 
  \begin{cases}
   \sfd_{BH,\Omega}^{(+)}
     &, \text{in BH-side} \\
   \int_{-\infty}^{\infty} d\omega \,
      A_{\Omega \omega}^{(1)} \, \sfd_{F , \omega}^{(1)}
     &, \text{in FRW-side} \, ,
  \end{cases}
  \label{eq-sec.sf.mf.TRWX-outTR.noref} \\
 \psi_{BH,\Omega}^{(-)} &=& 
  \begin{cases}
   \sfd_{BH,\Omega}^{(-)}
     &, \text{in BH-side} \\
   \int_{-\infty}^{\infty} d\omega \,
      B_{\Omega \omega}^{(2)} \, \sfd_{F , \omega}^{(2)}
     &, \text{in FRW-side} \, ,
  \end{cases}
  \label{eq-sec.sf.mf.TRWX-inTR.noref}
 \sikie
where $A^{(1)}$ and $B^{(2)}$ are summarized as
 \sikib
  A_{\Omega \omega}^{(1)}
   = \sgn(\omega) \,
      \frac{e^{-i \omega \rt_0}}{2\pi \sqrt{|\Omega \omega|}} \,
         I_A(\Omega,\omega)
   = \frac{e^{-i \omega \rt_0}}{2\pi} \,
      \sqrt{\left|\frac{\omega}{\Omega}\right|} \,
       \int_{-\infty}^{\infty} \, d\eta \,
         e^{i \omega \eta -i \, \Omega [T(\eta)-\Rt(\eta)]} \, ,
 \label{eq-sec.sf.mf.TRWX-coeff.A1}
 \sikie
and
 \sikibnon
    B_{\Omega \omega}^{(2)}
   = \frac{e^{i \omega \rt_0}}{2\pi} \,
      \sqrt{\left|\frac{\omega}{\Omega}\right|} \,
       \int_{-\infty}^{\infty} \, d\eta \,
         e^{i \omega \eta -i \, \Omega [T(\eta)+\Rt(\eta)]} \, .
 \sikienon
These results mean that the junction surface does not reflect the 
scalar field, therefore the out-going TR-mode is a superposition of 
the out-going modes in the FRW-side without including any in-going 
mode, and the in-going TR-mode consists of the in-going modes in the 
FRW-side. Further we can check by some routine computations that the 
inner products of these modes defined by 
(\ref{eq-sec.sf.mf.prep-inpro}) result in
 \sikib
  \inpro{\psi_{BH,\Omega}^{(\pm)}}{\psi_{BH,\Omega'}^{(\pm)}}
    = \sgn(\Omega) \, \delta(\Omega - \Omega') 
 \quad , \quad
    \inpro{\psi_{BH,\Omega}^{(\pm)}}{\psi_{BH,\Omega'}^{(\mp)}} =0
 \label{eq-sec.sf.mf.TRWX-inpro} \, .
 \sikie
This means that the out-going and in-going TR-modes given by 
(\ref{eq-sec.sf.mf.TRWX-outTR.noref}) and 
(\ref{eq-sec.sf.mf.TRWX-inTR.noref}) span a complete orthonormal 
basis. 

There are two reasons for the disappearance of reflection at $\js$: 
(i) the junction condition for the scalar field with no surface 
potential at $\js$, and (ii) the conformal flatness of the background 
spacetime. To understand the first reason, let us consider the case 
of $F \neq 0$ for the junction condition (\ref{eq-app.jc-jcsf.diff}) 
in the appendix \ref{app-jc}. In this case we can follow the same way 
of calculation to obtain a similar result to 
(\ref{eq-sec.sf.mf.TRWX-coeff.out}),
 \sikibnon
  A_{\Omega \omega}^{(1)} &=&
   (\text{the same form as (\ref{eq-sec.sf.mf.TRWX-coeff.out})})
  + \sgn(\omega) \, 2\pi \, i \, h_{F,\omega}^{(2)} \,
    \int_{-\infty}^{\infty} d\eta \,
      a \, f_{F,\omega}^{\ast} \, F  \\
  A_{\Omega \omega}^{(2)} &=&
   (\text{the same form as (\ref{eq-sec.sf.mf.TRWX-coeff.out})})
  - \sgn(\omega) \, 2\pi \, i \, h_{F,\omega}^{(1)} \,
    \int_{-\infty}^{\infty} d\eta \,
      a \, f_{F,\omega}^{\ast} \, F \, .
 \sikienon
The second term of $A^{(2)}$ does not vanish, and the reflection 
should take place at $\js$. On the other hand even if $F=0$, when the 
background spacetime is not conformal flat, $A^{(2)}$ cannot vanish 
due to the curvature scattering.

\subsubsection*{WX-mode}

The out-going and in-going WX-modes 
$\psiWX_{BH,\tilde{\Omega}}^{(\pm)}$ can be obtained by a similar way 
to the TR-mode. The resultant WX-mode is not reflected at $\js$ too, 
 \sikib
  \psiWX_{BH,\tilde{\Omega}}^{(+)} &=& 
  \begin{cases}
   \sfdWX_{BH,\tilde{\Omega}}^{(+)}
     &, \text{in BH-side} \\
   \int_{-\infty}^{\infty} d\omega \,
      \tilde{A}_{\tilde{\Omega} \omega}^{(1)} \,
        \sfd_{F , \omega}^{(1)}
     &, \text{in FRW-side} \, ,
  \end{cases}
  \label{eq-sec.sf.mf.TRWX-outWX.noref} \\
 \psiWX_{BH,\tilde{\Omega}}^{(-)} &=& 
  \begin{cases}
   \sfdWX_{BH,\tilde{\Omega}}^{(-)}
     &, \text{in BH-side} \\
   \int_{-\infty}^{\infty} d\omega \,
      \tilde{B}_{\tilde{\Omega} \omega}^{(2)} \,
        \sfd_{F , \omega}^{(2)}
     &, \text{in FRW-side} \, .
  \end{cases}
  \label{eq-sec.sf.mf.TRWX-inWX.noref}
 \sikie
The junction coefficients $\tilde{A}^{(1)}$ and $\tilde{B}^{(2)}$ are 
given by
 \sikibnon
  \tilde{A}_{\tilde{\Omega} \omega}^{(1)}
 &=& \frac{e^{-i\omega \rt_0}}{2\pi} \,
     \sqrt{\left| \frac{\omega}{\tilde{\Omega}} \right|} \,
     \int_{-\infty}^{\infty} d\eta \,
       e^{i\omega \eta - i\,\tilde{\Omega}[W(\eta)-X(\eta)]} \\
  \tilde{B}_{\tilde{\Omega} \omega}^{(2)}
 &=&  \frac{e^{i\omega \rt_0}}{2\pi} \,
     \sqrt{\left| \frac{\omega}{\tilde{\Omega}} \right|} \,
     \int_{-\infty}^{\infty} d\eta \,
       e^{i\omega \eta - i\,\tilde{\Omega}[W(\eta)+X(\eta)]} \, ,
 \sikienon
where $(W(\eta),X(\eta))$ is the coordinate of a point on $\js$ 
measured in the WX-system. These WX-modes 
$\psiWX_{BH,\tilde{\Omega}}^{(\pm)}$ also span a complete orthonormal 
basis.

\subsubsection{Out-going and in-going Cfl-modes}
\label{sec.sf.mf-Cfl}

The out-going and in-going Cfl-modes $\psi_{F,\omega}^{(1,2)}$ are 
monochromatic in the FRW-side and a superposition of some 
monochromatic modes in the BH-side. Because the reflection of $\SFD$ 
at $\js$ disappears as discussed in the previous subsection, these 
modes can be expressed as
 \sikib
  \psi_{F,\omega}^{(1)} &=& 
  \begin{cases}
   \int_{-\infty}^{\infty} d\Omega \,
      C_{\Omega \omega}^{(+)} \, \sfd_{BH,\Omega}^{(+)}
     &, \text{in BH-side} \\
   \sfd_{F,\omega}^{(1)}
     &, \text{in FRW-side} \, ,
  \end{cases}
  \label{eq-sec.sf.mf.Cfl-out.noref} \\
  \psi_{F,\omega}^{(2)} &=& 
  \begin{cases}
   \int_{-\infty}^{\infty} d\Omega \,
      D_{\Omega \omega}^{(-)} \, \sfd_{BH,\Omega}^{(-)}
     &, \text{in BH-side} \\
   \sfd_{F,\omega}^{(2)}
     &, \text{in FRW-side} \, .
  \end{cases}
  \label{eq-sec.sf.mf.Cfl-in.noref}
 \sikie
where we use the TR-mode in the BH-side for later use. The junction 
coefficients $C^{(+)}$ and $D^{(-)}$ can be determined in principle 
by the junction conditions of the scalar field. However because the 
junction surface does not rest in the TR-system 
$R(\eta) = a(\eta) \, r_0$, it is technically difficult to follow the 
same calculation done in the previous subsection. In other words, the 
completeness of the temporal part, $e^{-i\,\Omega T}$, cannot extract 
the simple integral representation of the junction coefficients 
$C^{(+)}$ and  $D^{(-)}$ like (\ref{eq-sec.sf.mf.TRWX-coeff.out}). 
However the analytic representations of them will be obtained in next 
section \ref{sec.sf-quant} with making use of the Bogoljubov 
transformation. Therefore we assume temporarily that these Cfl-modes 
span a complete orthonormal basis. After obtaining the integral 
representations of $C^{(+)}$ and $D^{(-)}$, we will find that this 
assumption is consistent. 

We can easily find some relations between $A^{(1)}$, $B^{(2)}$ and 
$C^{(+)}$, $D^{(-)}$. By substituting 
(\ref{eq-sec.sf.mf.TRWX-outTR.noref}) into 
(\ref{eq-sec.sf.mf.Cfl-out.noref}) and other similar operations, we 
obtain
 \begin{align}
   \int_{-\infty}^{\infty} d\omega \,
       A_{\Omega,\omega}^{(1)} \, C_{\Omega' \omega}^{(+)}
    &= \delta\left( \Omega - \Omega' \right) 
 & \int_{-\infty}^{\infty} d\Omega \,
       A_{\Omega,\omega}^{(1)} \, C_{\Omega \omega'}^{(+)}
    &= \delta\left( \omega - \omega' \right)
 \label{eq-sec.sf.mf.Cfl-coeff.rel.1} \\
   \int_{-\infty}^{\infty} d\omega \,
       B_{\Omega,\omega}^{(2)} \, D_{\Omega' \omega}^{(-)}
    &= \delta\left( \Omega - \Omega' \right) 
 & \int_{-\infty}^{\infty} d\Omega \,
       B_{\Omega,\omega}^{(2)} \, D_{\Omega \omega'}^{(-)}
    &= \delta\left( \omega - \omega' \right)
  \label{eq-sec.sf.mf.Cfl-coeff.rel.2} \, .
 \end{align}

\subsection{Quantization}
\label{sec.sf-quant}

\subsubsection{Vacuum states of WX-mode, TR-mode and Cfl-mode}
\label{sec.sf.quant-vac}

\subsubsection*{Positive frequency modes}

For canonical quantization of the scalar field $\SFD$, we need the 
positive frequency mode. Hereafter as the notation, we denotes the 
positive frequency TR-mode, WX-mode and Cfl-mode respectively as 
$\Psi_{BH}$, $\PsiWX_{BH}$ and $\Psi_{F}$. These modes can be 
obtained from the out-going and in-going modes as
 \sikibnon
  \Psi_{BH,K} =
   \begin{cases}
    \psi_{BH,\Omega=K}^{(+)} &, \text{for $K>0$} \\
    \psi_{BH,\Omega=-K}^{(-)} &, \text{for $K<0$} \, ,
   \end{cases}
  \quad \text{(Positive Frequency TR-mode)}
 \sikienon
where $\Omega$ is the frequency, $K$ is the wave number and 
$\Omega=|K|$. Similarly we find
 \sikibnon
  \PsiWX_{BH,K} =
   \begin{cases}
    \psi_{BH,\tilde{\Omega}=\tilde{K}}^{(+)}
      &, \text{for $\tilde{K}>0$} \\
    \psi_{BH,\tilde{\Omega}=-\tilde{K}}^{(-)}
      &, \text{for $\tilde{K}<0$} \, ,
   \end{cases}
  \quad \text{(Positive Frequency WX-mode)}
 \sikienon
and
 \sikibnon
  \Psi_{F,k} =
   \begin{cases}
    \psi_{F,\omega=k}^{(1)} &, \text{for $k>0$} \\
    \psi_{F,\omega=-k}^{(2)} &, \text{for $k<0$} \, ,
   \end{cases}
  \quad \text{(Positive Frequency Cfl-mode)}
 \sikienon
where $\tilde{\Omega}=|\tilde{K}|$ and $\omega=|k|$. For example, the 
positive frequency TR-mode in the BH-side is expressed as
 \sikibnon
  \Psi_{BH,K} \, \text{in BH-side}
 = \frac{1}{\sqrt{4\pi|K|}} \, e^{-i(\Omega T - K \Rt)}
 = \frac{1}{\sqrt{4\pi|K|}} \times
  \begin{cases}
    e^{-i K (T - \Rt)} &, \text{for $K>0$} \\
    e^{+i K (T + \Rt)} &, \text{for $K<0$} \, ,
  \end{cases}
 \sikienon
and this satisfies the definition of the positive frequency mode 
${\cal L}_{\xi_{BH}} \Psi_{BH} = -i \Omega \Psi_{BH}$ where 
$\Omega > 0$.

We should note that the mode functions $\Psi_{BH}$ and $\Psi_{F}$ are 
defined only in the R-region or in the L-region of the maximally 
extended eternal SCU. Therefore we define the other mode functions as 
\cite{ref-unruh}
 \sikibnon
  _R\Psi_{BH,K} =
    \begin{cases}
     \Psi_{BH,K} &, \text{in R-region} \\
     0           &, \text{in L-region}
    \end{cases}
 \quad , \quad
  _L\Psi_{BH,K} =
    \begin{cases}
     0           &, \text{in R-region} \\
     \Psi_{BH,K}^{\ast} &, \text{in L-region} \, ,
    \end{cases}
 \sikienon
and
 \sikibnon
  _R\Psi_{F,k} =
    \begin{cases}
     \Psi_{F,k} &, \text{in R-region} \\
     0           &, \text{in L-region}
    \end{cases}
 \quad , \quad
  _L\Psi_{F,k} =
    \begin{cases}
     0           &, \text{in R-region} \\
     \Psi_{F,k}^{\ast} &, \text{in L-region} \, .
    \end{cases}
 \sikienon
All of these modes are of the positive frequency, and can be defined 
on a Cauchy surface connecting the spatial infinities of R-region and 
L-region via the bifurcation point of the event horizon. We take the 
complex conjugate in the definition of $_L\Psi_{BH,K}$ and 
$_L\Psi_{F,k}$, since the future temporal direction in the L-region 
with respect to $T$ and $\eta$ is inverted in comparison with that in 
the R-region. Further it is useful to define the following mode 
functions \cite{ref-unruh}, 
 \begin{equation}
 \begin{split}
  \PsiWXbyTR_{BH,K}^{I}
 \,\,\, &= \,\,\,
    \frac{1}{|2\sinh(\pi\Omega/\sg)|^{1/2}} \,
     \left[ \, e^{\pi\Omega/2\sg} \, _R\Psi_{BH,K}
             + e^{-\pi\Omega/2\sg} \, _L\Psi_{BH,K}^{\ast} \,
     \right] \\
  \PsiWXbyTR_{BH,K}^{II}
 \,\,\, &= \,\,\,
    \frac{1}{|2\sinh(\pi\Omega/\sg)|^{1/2}} \,
     \left[ \, e^{-\pi\Omega/2\sg} \, _R\Psi_{BH,K}^{\ast}
          + e^{\pi\Omega/2\sg} \, _L\Psi_{BH,K} \,
     \right] \, .
 \end{split}
 \label{eq-sec.sf.quant.vac-bogo.WXtoTR}
 \end{equation}
These mode functions share the same analyticity as the mode 
$\PsiWX_{BH}$. That is, $\PsiWXbyTR_{BH}^{I\, , \,II}$ and 
$\PsiWX_{BH}$ are analytic and bounded on the lower-half-plane of 
the complex $U (= W \pm X)$ plane. 

Due to the completeness of the out-going and in-going TR-modes 
(\ref{eq-sec.sf.mf.TRWX-inpro}), the positive frequency TR-modes 
satisfy the complete orthonormal relations. So is the positive 
frequency WX-modes. For the positive frequency Cfl-modes, the 
assumption mentioned at the subsection \ref{sec.sf.mf-Cfl} makes the 
positive frequency modes satisfy the complete orthonormal relations. 
Hence we can carry out the canonical quantization of $\SFD$ using 
each mode function.

\subsubsection*{Quantization with WX-mode}

The scalar field $\SFD$ is quantized using the WX-mode as
 \sikibnon
  \SFD = \int_{-\infty}^{\infty} d\tilde{K} \, 
     \left[ \tilde{a}_{\tilde{K}} \, \PsiWX_{BH,\tilde{K}}
          + \tilde{a}_{\tilde{K}}^{\dag} \,
             \PsiWX_{BH,\tilde{K}}^{\ast} \right] \, .
 \sikienon
Further because $\PsiWXbyTR_{BH}$ and $\PsiWX_{BH}$ share the same 
analyticity, we can find another representation,
 \sikibnon
  \SFD = \int_{-\infty}^{\infty} dK \, 
     \left[ a_K^{I} \, \PsiWXbyTR_{BH,K}^{I}
          + a_K^{I\,\dag} \, \PsiWXbyTR_{BH,K}^{I\,\ast}
          + a_K^{II} \, \PsiWXbyTR_{BH,K}^{II}
          + a_K^{II\,\dag} \, \PsiWXbyTR_{BH,K}^{II\,\ast}
     \right] \, .
 \sikienon
Here $\tilde{a}_{\tilde{K}}$ and $a_K^{I,II}$ are the annihilation 
operators, and their Hermitian conjugates are the creation operators. 
Due to the same analyticity of $\PsiWXbyTR_{BH}^{I\,,\,II}$ as 
$\PsiWX_{BH}$, they also share the same vacuum state $\vac_W$ defined 
by
 \sikibnon
  \tilde{a}_{\tilde{K}} \, \vac_W
 = a_K^I  \, \vac_W = a_K^{II}  \, \vac_W = 0
 \quad \text{for all $\tilde{K}$ and $K$} \, .
 \sikienon

\subsubsection*{Quantization with TR-mode}

The quantization of $\SFD$ using the TR-mode is represented as
 \sikibnon
  \SFD = \int_{-\infty}^{\infty} dK \, 
    \left[ \, _Rb_K \, _R\Psi_{BH,K}
         + \, _Rb_K^{\dag} \, _R\Psi_{BH,K}^{\ast}
         + \, _Lb_K \, _L\Psi_{BH,K}
         + \, _Lb_K^{\dag} \, _L\Psi_{BH,K}^{\ast} \,
    \right] \, ,
 \sikienon
where $_Rb_K$ and $_Lb_K$ are the annihilation operators and 
$_Rb_K^{\dag}$ and $_Lb_K^{\dag}$ are the creation ones. The vacuum 
state of the TR-mode $\vac_T$ is defined by
 \sikibnon
  _Rb_K \, \vac_T = _Lb_K \, \vac_T = 0
   \quad \text{for all $K$} \, .
 \sikienon

\subsubsection*{Quantization with Cfl-mode}

The quantization with the Cfl-mode is similar to that of the 
TR-mode,
 \sikibnon
  \SFD = \int_{-\infty}^{\infty} dK \, 
    \left[ \, _Rc_k \, _R\Psi_{F,k}
         + \, _Rc_k^{\dag} \, _R\Psi_{F,k}^{\ast}
         + \, _Lc_k \, _L\Psi_{BH,K}
         + \, _Lc_k^{\dag} \, _L\Psi_{F,k}^{\ast} \,
    \right] \, ,
 \sikienon
where $_Rc_k$ and $_Lc_k$ are the annihilation operators and 
$_Rc_k^{\dag}$ and $_Lc_k^{\dag}$ are the creation ones. The vacuum 
state of the Cfl-mode $\vac_{\eta}$ is defined by
 \sikibnon
  _Rc_k \, \vac_{\eta} = _Lc_k \, \vac_{\eta} = 0
   \quad \text{for all $k$} \, .
 \sikienon

\subsubsection{Bogoljubov transformation}
\label{sec.sf.quant-bogo}

The Bogoljubov transformation is the change of positive frequency 
basis. Therefore the equation (\ref{eq-sec.sf.quant.vac-bogo.WXtoTR}) 
gives the Bogoljubov transformation between the WX-mode and the 
TR-mode. In this subsection, we derive the Bogoljubov 
transformation between the TR-mode and the Cfl-mode. It is important 
to notice the implication of the relations 
(\ref{eq-sec.sf.mf.Cfl-coeff.rel.1}) and 
(\ref{eq-sec.sf.mf.Cfl-coeff.rel.2}) that, for the out-going modes,
 \sikibnon
  \int_{-\infty}^{\infty} d\Omega \,
    C_{\Omega \omega}^{(+)} \, \psi_{BH,\Omega}^{(+)}
 &=& \int_{-\infty}^{\infty} d\Omega \,
    C_{\Omega \omega}^{(+)} \, \sfd_{BH,\Omega}^{(+)}
 = \psi_{F,\omega}^{(1)}
  \quad \text{in BH-side} \\
 &=& \iint_{-\infty}^{\infty} d\omega' \, d\Omega \,
     C_{\Omega \omega}^{(+)} \, A_{\Omega \omega'} \,
      \sfd_{F,\omega'}^{(1)}
 = \psi_{F,\omega}^{(1)}
   \quad \text{in FRW-side} \, ,
 \sikienon
and a similar one for the in-going mode. This denotes that the 
junction coefficients $A^{(1)}$, $B^{(2)}$, $C^{(+)}$ and $D^{(-)}$ 
give the change of out-going and in-going basis as
 \sikibnon
  \psi_{BH,\Omega}^{(+)}
 = \int_{-\infty}^{\infty} d\omega \,
      A_{\Omega \omega}^{(1)} \, \psi_{F,\omega}^{(1)}
 \quad , \quad
  \psi_{BH,\Omega}^{(-)}
 = \int_{-\infty}^{\infty} d\omega \,
      B_{\Omega \omega}^{(2)} \, \psi_{F,\omega}^{(2)} \, ,
 \sikienon
and
 \sikibnon
  \psi_{F,\omega}^{(1)}
 = \int_{-\infty}^{\infty} d\omega \,
      C_{\Omega \omega}^{(+)} \, \psi_{BH,\Omega}^{(+)}
 \quad , \quad
  \psi_{F,\omega}^{(2)}
 = \int_{-\infty}^{\infty} d\omega \,
      D_{\Omega \omega}^{(-)} \, \psi_{BH,\Omega}^{(-)} \, .
 \sikienon
Consequently, by definition of $\Psi_{BH}$, $\Psi_F$, 
$\psi_{BH}^{(\pm)}$ and $\psi_F^{(1,2)}$, we obtain the Bogoljubov 
transformation between the positive frequency modes $\Psi_{BH}$ and 
$\Psi_F$ as
 \sikibnon
  \Psi_{BH,K} &=& \int_{-\infty}^{\infty} dk \,
    \left[ \alpha_{K k} \, \Psi_{F,k}
         + \beta_{K k} \, \Psi_{F,k}^{\ast} \right] \\
  \Psi_{F,k} &=& \int_{-\infty}^{\infty} dK \,
    \left[ \lambda_{K k} \, \Psi_{BH,K}
         + \mu_{K k} \, \Psi_{BH,K}^{\ast} \right] \, ,
 \sikienon
where the Bogoljubov coefficients are given as
 \begin{equation}
 \begin{split}
  \alpha_{K k} \,\,\, &= \,\,\,
   \begin{cases}
    A_{K k}^{(1)} &, \text{for $K>0$, $k>0$} \\
    0 &, \text{for $K>0$ and $k<0$, or $K<0$ and $k>0$} \\
    B_{-K -k}^{(2)} &, \text{for $K<0$, $k<0$} \, ,
   \end{cases}
 \\
  \beta_{K k} \,\,\, &= \,\,\,
   \begin{cases}
    A_{K -k}^{(1)} &, \text{for $K>0$, $k>0$} \\
    0 &, \text{for $K>0$ and $k<0$, or $K<0$ and $k>0$} \\
    B_{-K k}^{(2)} &, \text{for $K<0$, $k<0$} \, ,
   \end{cases}
 \label{eq-sec.sf.quant.bogo-coeff}
 \end{split}
 \end{equation}
and
 \begin{equation*}
 \begin{split}
  \lambda_{K k} \,\,\, &= \,\,\,
   \begin{cases}
    C_{K k}^{(+)} &, \text{for $K>0$, $k>0$} \\
    0 &, \text{for $K>0$ and $k<0$, or $K<0$ and $k>0$} \\
    D_{-K -k}^{(-)} &, \text{for $K<0$, $k<0$} \, ,
   \end{cases}
 \\
  \mu_{K k} \,\,\, &= \,\,\,
   \begin{cases}
    C_{-K k}^{(+)} &, \text{for $K>0$, $k>0$} \\
    0 &, \text{for $K>0$ and $k<0$, or $K<0$ and $k>0$} \\
    D_{K -k}^{(-)} &, \text{for $K<0$, $k<0$} \, .
   \end{cases}
 \end{split}
 \end{equation*}
Further, taking the inverse of the Bogoljubov transformation, we can 
find the relation between $(\alpha , \beta)$ and $(\lambda , \mu)$ to 
be $\lambda_{K k} = \alpha_{K k}^{\ast}$, 
$\mu_{K k} = - \beta_{K k}^{\ast}$. Hence the Bogoljubov 
transformation from $(_R\Psi_{BH,K} \, , \, _L\Psi_{BH,K})$ to 
$(_R\Psi_{F,k} \, , \, _L\Psi_{F,k})$ is obtained to be
 \begin{equation}
 \begin{split}
  _R\Psi_{BH,K} \,\,\, &= \,\,\,
    \int_{-\infty}^{\infty} dk \,
      \left[\, \alpha_{K k} \,\, _R\Psi_{F,k}
             + \beta_{K k} \,\, _R\Psi_{F,k}^{\ast} \, \right] \\
  _L\Psi_{BH,K} \,\,\, &= \,\,\,
    \int_{-\infty}^{\infty} dk \,
      \left[\, \alpha_{K k}^{\ast} \,\, _L\Psi_{F,k}
             + \beta_{K k}^{\ast} \,\, _L\Psi_{F,k}^{\ast} \,
      \right]
 \label{eq-sec.sf.quant.bogo-bogo.TRtoCfl} \, ,
 \end{split}
 \end{equation}

Finally note that the junction coefficients $C^{(+)}$ and $D^{(-)}$ 
are represented analytically using the other junction coefficients 
$A^{(1)}$ and $B^{(2)}$ through the Bogoljubov transformations 
obtained above. Therefore we can find that the out-going and in-going 
Cfl-modes span a complete orthonormal basis.


\section{Hawking radiation in the SCU}
\label{sec-hr}

\subsection{Hawking radiation}
\label{sec.hr-hr}

We compute the Hawking radiation observed by a comoving observer. 
As discussed at the section \ref{sec.scu-str}, its spectrum is given 
by $_W\dvac N_{F,\omega} \vac_W$, where 
$N_{F,\omega} = \, _Rc_{\omega}^{\dag} \, _Rc_{\omega}$ is the number 
operator of particles of the Cfl-mode, 
\footnote{Here we set the observer in the R-region. But even if the 
observer is in the L-region, the resultant spectrum is the same.}
and $\omega (>0)$ is the energy of the particle which is equivalent 
to the frequency of the positive frequency Cfl-mode. Further in order 
to treat the situation that the observer receives the out-going 
radiation from the black hole, we should consider only the out-going 
Cfl-mode, $k>0$. Then taking the Bogoljubov transformations 
(\ref{eq-sec.sf.quant.vac-bogo.WXtoTR}) and 
(\ref{eq-sec.sf.quant.bogo-bogo.TRtoCfl}) successively, the Hawking 
spectrum is obtained to be
 \sikib
  _W\dvac N_{F,\omega} \vac_W = \int_{0}^{\infty} d\Omega \,
    \frac{1}{e^{2\pi\Omega/\sg}-1} \, D_H(\Omega,\omega) \, ,
 \label{eq-sec.hr.hr-hr}
 \sikie
where $D_H$ is given by
 \sikib
  D_H(\Omega,\omega) =
   \left| A_{\Omega\, \omega}^{(1)} \right|^2 
 + e^{2\pi\Omega/\sg} \, \left| A_{\Omega\, -\omega}^{(1)} \right|^2 
 + 2 \, e^{\pi\Omega/\sg} \,
        \Re\left( A_{\Omega\, \omega}^{(1)} \,
                  A_{\Omega\, -\omega}^{(1)} \right) \, ,
 \label{eq-sec.hr.hr-dev}
 \sikie
and $\Re$ denotes the real part, $A^{(1)}$ is given at 
(\ref{eq-sec.sf.mf.TRWX-coeff.A1}), the relations $\Omega=|K|=K$ and 
$\omega=|k|=k$ are used, and due to the relations 
(\ref{eq-sec.sf.quant.bogo-coeff}) the variable of integration is 
restricted to $K>0 \Rightarrow \Omega>0$.

Obviously the Hawking radiation in the SCU is totally different from 
a thermal spectrum.
\footnote{But the case that $a=$ constant should reproduce the 
thermal spectrum.}
The factor $D_H$ represents such a difference, and we call $D_H$ the 
{\it deviation factor}. Note that the particle creation due to the 
cosmological expansion does not take place with a minimal coupling 
massless scalar field in two dimensional background spacetime 
\cite{ref-bd}. So the deviation factor $D_H$ includes only the effect 
of the motion of the observer relative to the TR-system, and does not 
include the cosmological particle creation. 

The other notice of the spectrum (\ref{eq-sec.hr.hr-hr}) is that 
the Bogoljubov transformations used in deriving the Hawking radiation 
are defined all over the spacetime. Especially whole of the temporal 
information from $\eta=-\infty$ to $\eta=\infty$ are included in this 
result. That is, we should interpret the observer is at a remote 
future $\eta \to +\infty$.

\subsection{Divergence and normalization of the Hawking spectrum}
\label{sec.hr-dn}

The black hole should lose its mass energy $M$ due to the Hawking 
radiation. However, since the back reaction to the background 
spacetime is ignored in our derivation of the resultant spectrum 
(\ref{eq-sec.hr.hr-hr}), the mass $M$ remains constant during 
emitting the Hawking radiation. Thus, the spectrum 
(\ref{eq-sec.hr.hr-hr}) may give us the conclusion that the observer 
at a remote future receives an infinite energy from the black hole 
via the Hawking radiation. In the next subsection 
\ref{sec.hr.dn-div}, we see the divergence of the spectrum 
(\ref{eq-sec.hr.hr-hr}) for the simplest case that the scale factor 
is constant. Then, the normalization of the spectrum 
(\ref{eq-sec.hr.hr-hr}) is discussed in the following subsection 
\ref{sec.hr.dn-norm}.

\subsubsection{Simplest case: constant scale factor}
\label{sec.hr.dn-div}

Consider the case $a = a_c =$ constant, which should give the thermal 
spectrum. From (\ref{eq-sec.scu.scu-js}), $R(\eta) = a_c r_0 =$ 
constant, then (\ref{eq-sec.scu.scu-time}) gives 
$T(\eta) = b_c \, \eta \, +$ constant, where 
$b_c = a_c \sqrt{1-k \, r_0^{\,\,2}}/(1-R_g/a_c r_0)$. That is, 
$T(\eta) - \Rt(\eta) = b_c \, \eta - \Rt_c$, where $\Rt_c =$ 
constant. Then the junction coefficient $A_{\Omega \omega}^{(1)}$ 
given at (\ref{eq-sec.sf.mf.TRWX-coeff.A1}) is calculated using the 
representation of the delta function 
$\int_{-\infty}^{\infty}dk \exp(i k x) = 2\pi \delta(x)$,
 \sikib
  A_{\Omega \omega}^{(1)} =
   \sqrt{\left|\frac{\omega}{\Omega}\right|} \, 
   e^{-i (\omega \rt_0 - \Omega \Rt_c)} \,
   \delta(\omega - b_c \Omega) \, .
 \label{eq-sec.hr.dn.div-A1}
 \sikie
Since $A_{\Omega\, -\omega}^{(1)}=0$ for $\Omega>0$ and $\omega>0$, 
the deviation factor becomes $D_H = |A_{\Omega \omega}^{(1)}|^2$ 
which gives the spectrum (\ref{eq-sec.hr.hr-hr}) as,
 \sikib
  _W\dvac N_{F,\omega} \vac_W =
    \frac{b_c}{e^{2\pi\omega/b_c \sg}-1} \, \delta(0) \, .
 \label{eq-sec.hr.dn.div-hr}
 \sikie
This spectrum diverges by the delta function $\delta(0)$.

\subsubsection{Finite time interval normalization}
\label{sec.hr.dn-norm}

If the delta function appeared in (\ref{eq-sec.hr.dn.div-A1}) is 
removed, the divergence in the previous resultant spectrum 
(\ref{eq-sec.hr.dn.div-hr}) is normalized. 
Note that the delta function in (\ref{eq-sec.hr.dn.div-A1}) comes 
from the integral of infinite time interval in the representation 
(\ref{eq-sec.sf.mf.TRWX-coeff.A1}). Therefore we make the comoving 
observer receive the Hawking radiation during a finite time interval, 
$\eta_i < \eta < \eta_f (= \eta_i + \eta_p)$. Further we introduce 
the periodic boundary condition for the scalar field, 
$\SFD(\eta + \eta_p) = \SFD(\eta)$, which changes the frequency of 
the Cfl-mode $\omega$ and that of the TR-mode $\Omega$ with the 
discrete quantities,
 \begin{equation}
  \begin{split}
    \omega = \frac{2\pi}{\eta_p} \, n
     \quad &, \quad n = \pm1, \pm2, \pm3, \cdots
       \quad \text{and} \quad \eta_p = \eta_f - \eta_i \\
    \Omega = \frac{2\pi}{T_p} \, N 
     \quad &, \quad N = \pm1, \pm2, \pm3, \cdots
       \quad \text{and} \quad T_p = T(\eta_f) - T(\eta_i) \, ,
  \end{split}
 \label{eq-sec.hr.dn.norm-freq}
 \end{equation}
where, because the mode solution of zero frequency is trivial 
$\SFD =$ constant, the zero frequencies $n=0$ and $N=0$ are 
neglected. The integral with respect to $\omega$, $\Omega$ 
and $\eta$ should be replaced as
 \sikibnon
  \int_{-\infty}^{\infty} d\omega \, \longrightarrow \,
    \sum_{n=-\infty \, , \, \neq 0}^{\infty}
 \quad , \quad
  \int_{-\infty}^{\infty} d\Omega \, \longrightarrow \,
    \sum_{N=-\infty \, , \, \neq 0}^{\infty}
 \quad , \quad
  \int_{-\infty}^{\infty} d\eta \, \longrightarrow \,
    \int_{\eta_i}^{\eta_f} d\eta \, ,
 \sikienon
and one of the equation in the completeness 
(\ref{eq-sec.sf.mf.prep-Cfl.comp}) is changed to the form
$\int_{\eta_i}^{\eta_f} d\eta \,
  f_{F,\omega}^{\ast}(\eta) \, f_{F,\omega'}(\eta)
    = \eta_p \, \delta_{\omega , \omega'}$. Consequently, the 
junction coefficient $A_{\Omega \omega}^{(1)}$ can be calculated by 
the similar way to obtain (\ref{eq-sec.sf.mf.TRWX-coeff.A1}),
 \sikib
  A_{\Omega \omega}^{(1)}
   = \frac{e^{-i \omega \rt_0}}{\eta_p} \,
      \sqrt{\left|\frac{\omega}{\Omega}\right|} \,
       \int_{\eta_i}^{\eta_f} \, d\eta \,
         e^{i \omega \eta -i \, \Omega [T(\eta)-\Rt(\eta)]} \, ,
 \label{eq-sec.hr.dn.norm-A1}
 \sikie
and $A_{\Omega \omega}^{(2)}=0$ should also hold. Finally the Hawking 
spectrum (\ref{eq-sec.hr.hr-hr}) is modified to the following form, 
 \sikib
  I_H(\eta_i , \eta_f ; \omega) =
    \sum_{N=1}^{\infty} \,
      \frac{1}{e^{2\pi\Omega/\kappa}-1} \, D_H(\Omega,\omega) \, ,
 \label{eq-sec.hr.dn.norm-hr}
 \sikie
where $\omega$ and $\Omega$ are given by 
(\ref{eq-sec.hr.dn.norm-freq}), and the deviation factor 
$D_H(\Omega , \omega)$ is given by (\ref{eq-sec.hr.hr-dev}). 

Applying this finite time interval normalization to the simplest case 
of the previous subsection \ref{sec.hr.dn-div}, the delta function in 
(\ref{eq-sec.hr.dn.div-A1}) is replaced by the Kronecker delta 
$\delta_{\omega, b_c \Omega}$. Then the spectrum 
(\ref{eq-sec.hr.dn.div-hr}) is normalized to the form 
(\ref{eq-sec.hr.dn.norm-hr}), 
$I_H(\eta_i , \eta_f; \omega)
  = b_c (e^{2\pi\omega/b_c \kappa}-1)^{-1}$.

\subsection{Some models}
\label{sec.hr-model}

In order to understand the properties of the normalized 
Hawking spectrum in the SCU (\ref{eq-sec.hr.dn.norm-hr}), it is 
useful to consider some cases of the phase factor, 
$T(\eta) - \Rt(\eta)$, in the representation 
(\ref{eq-sec.hr.dn.norm-A1}).

\subsubsection{Case: the thermal spectrum}
\label{sec.hr.model-therm}

Here we look for the scale factor $a(\eta)$ which makes the spectrum 
(\ref{eq-sec.hr.dn.norm-hr}) be the thermal one. According to the 
previous section \ref{sec.hr-dn}, the phase factor, 
$T(\eta) - \Rt(\eta)$, should be linear of $\eta$,
 \sikib
  T(\eta) - \Rt(\eta) = b \, \eta + p
 \quad , \quad b = \text{const} >0 \quad , \quad p = \text{const}. 
 \label{eq-sec.hr.model.therm-phase}
 \sikie
The spectrum (\ref{eq-sec.hr.dn.norm-hr}) of this case becomes 
thermal, 
$I_H(\eta_i , \eta_f; \omega) = b (e^{2\pi\omega/b \kappa}-1)^{-1}$. 
The differential of the phase factor 
(\ref{eq-sec.hr.model.therm-phase}) gives the following, 
 \sikib
  \od{\,a(\eta)}{\eta} =
    \frac{\sqrt{1-k\,r_0^{\,\,2}}}{r_0} \, a(\eta)
  - \frac{b}{r_0} \,
     \left( 1- \frac{R_g}{a(\eta) \, r_0} \right) \, .
 \label{eq-sec.hr.model.therm-sfac}
 \sikie
Though the simplest case, $a=$ constant, is the trivial solution of 
this equation, the general solution is a dynamical universe, $a \ne$ 
constant. 

For sufficiently small $b$, we can set $da/d\eta|_{\eta_0} > 0$ at 
an initial time $\eta_0$, and generally the scale factor grows 
exponentially. If we set roughly $a(\eta) = e^{\alpha \eta}$, where 
$\alpha$ is a constant, then it is rewritten to the form 
$a(t) \propto t$ through the relation $dt = a(\eta)\,d\eta$. This is 
the asymptotic form of the scale factor for the open FRW universe at 
a remote future $t \to +\infty$. That is, one may think that the 
black hole in the open universe emits a thermal Hawking radiation. 
However the situation is not so simple, as discussed in the following 
two subsections.

\subsubsection{Case: $T(\eta) - \Rt(\eta) \propto \eta^2$}
\label{sec.hr.model-to2}

In this subsection we treat a case that the scale factor gives 
the phase factor of the junction coefficient 
(\ref{eq-sec.hr.dn.norm-A1}) as 
 \sikibnon
  T(\eta) - \Rt(\eta) = c \, \eta^2 + b \, \eta + p
 \quad , \quad
  c = \text{const} >0 \quad , \quad b, p = \text{const}.
 \sikienon
Taking the differential of this equation, we find
 \sikib
  \od{\,a(\eta)}{\eta} =
    \frac{\sqrt{1-k\,r_0^{\,\,2}}}{r_0} \, a(\eta)
  - \frac{2\, c \, \eta + b}{r_0} \,
     \left( 1- \frac{R_g}{a(\eta) \, r_0} \right) \, .
 \label{eq-sec.hr.model.to2-sfac}
 \sikie
Generally for the positive initial value $a(\eta_0)>0$ at an initial 
time $\eta_0<-b/2c$, this equation makes the scale factor $a(\eta)$ 
grow exponentially. Therefore by the same discussion given in the 
previous case, the present case also includes the asymptotic form of 
the open FRW universe at a remote future. Though the rough estimates 
of the equations (\ref{eq-sec.hr.model.therm-sfac}) and 
(\ref{eq-sec.hr.model.to2-sfac}) gives the same conclusion that 
$a(t) \propto t$, but the difference of them is the detailed behavior 
of $a(t)$. For example, because $c>0$, the velocity of the expansion 
$da/d\eta$ of the present case is smaller than that of the previous 
case for $\eta > 0$. 

\begin{figure}[t]
  \cenb
   \includegraphics[height=25mm]{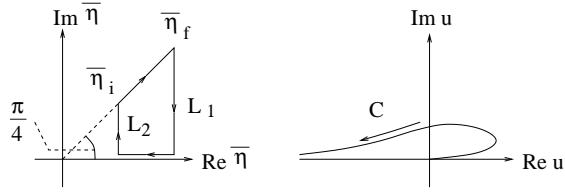}
  \cene
 \caption{{\small The left figure shows the integration path for 
(\ref{eq-sec.hr.model.to2-int}) in the $\eb$-plane, and the right one 
for (\ref{eq-sec.hr.model.ds-jc}) in the $u$-plane.}}
 \label{fig-path}
 \end{figure}

The spectrum (\ref{eq-sec.hr.dn.norm-hr}) of the present case is 
expected not to be the thermal one, since the phase factor, 
$T(\eta) - \Rt(\eta)$, is different from that of the previous case. 
The normalized junction coefficient (\ref{eq-sec.hr.dn.norm-A1}) is 
rewritten using the change of the conformal time, 
$\eb = \sqrt{c \, \Omega} \,
              e^{i\pi/4} \, [\eta - (\omega - b\Omega)/2c\Omega]$, 
 \sikib
  A_{\Omega \omega}^{(1)} =
   \frac{e^{-i p_c} }{\eta_p \, \Omega} \, 
   \sqrt{\frac{|\omega|}{c}} \,
   \int_{\eb_i}^{\eb_f} d\eb \, e^{-\eb^2} \, ,
 \label{eq-sec.hr.model.to2-junc.coeff}
 \sikie
where we set $\Omega (=2\pi N/T_p) >0$, that is, $N = 1,2,3,\cdots$, 
and 
$p_c =\omega\rt_0 -i(\omega-b\Omega)^2/4c\,\Omega +p\,\Omega +\pi/4$. 
Because the integrand $e^{-\eb^2}$ has no pole on the complex 
$\eb$-plane, the integral in this representation of 
$A^{(1)}$ becomes
 \sikib
  \int_{\eb_i}^{\eb_f}
  = - \int_{L_1} - \int_{L_2}
    - \int_{\Re(\eb_f)}^{\Re(\eb_i)} \, ,
 \label{eq-sec.hr.model.to2-int}
 \sikie
where the integration paths are shown at the figure \ref{fig-path}. 
Note that the path $L_2$ is given by $\eb=\Re(\eb_i) + i y$ for 
$0<y<\Im(\eb_i)$ where $\Im$ denotes the imaginary part, then we find 
 \sikibnon
  \int_{L_2} d\eb \, e^{-\eb^2}
 = i e^{-\Re(\eb_i)^2} \, \int_0^{\Im(\eb_i)} dy \,
       e^{y^2 - 2i \Re(\eb_i) y} 
 \longrightarrow 0 \quad \text{as $|\eta_i| \to \infty$} \, ,
 \sikienon
where 
$\Re(\eb_i)^2 =
   (c\,\Omega/2) \, [\eta_i - (\omega-b\,\Omega)/2c\,\Omega]^2$. 
Similarly it is found that $\int_{L_1} \to 0$ as 
$|\eta_i| \to \infty$. Therefore we obtain the approximate form 
of $A^{(1)}$ for sufficiently large $\eta_i$, 
 \sikibnon
  A_{\Omega \omega}^{(1)} \simeq
    \frac{e^{-i p_c} }{\eta_p \, \Omega} \, 
     \sqrt{\frac{|\omega|}{c}} \,
       \int_{\Re(\eb_i)}^{\Re(\eb_f)} d\eb \, e^{-\eb^2}
 =  \frac{e^{-i p_c} }{\eta_p \, \Omega} \, 
     \sqrt{\frac{|\omega|}{c}} \,
       \left[ \, \E( \Re(\eb_i) ) - \E( \Re(\eb_f) ) \, \right] \, ,
 \sikienon
where $\E(z) = \int_z^{\infty} du \, e^{-u^2}$ is the Gauss's error 
function. This gives the deviation factor (\ref{eq-sec.hr.hr-dev}) 
for large $\eta_i$, 
 \sikib
  D_H(\Omega , \omega) &\simeq&
   \frac{\omega}{c\,\Omega^2 \eta_p^{\,\,2}} \, 
   \left[
      \DE(\Omega,\omega)^2
    + e^{2\pi\Omega/\kappa} \, \DE(\Omega,-\omega)^2
 \right. \nonumber \\ && \qquad \qquad \left.
    + 2e^{\pi\Omega/\kappa} \,
       \cos\left( \frac{b^2\Omega^2+\omega^2}{4c\,\Omega}
                 -2p\,\Omega -\frac{\pi}{2} \right) \,
       \DE(\Omega,\omega) \, \DE(\Omega,-\omega)
   \right] \, ,
 \label{eq-sec.hr.model.to2-dev}
 \sikie
where $\DE(\Omega,\omega)=\E( \Re(\eb_i) ) -\E( \Re(\eb_f) )$, 
$\Omega=2\pi N/T_p$, $\omega=2\pi n/\eta_p$, and 
$N, n = 1,2,3,\cdots$. Note that $\Re(\eb)$ depends on $\Omega$ and 
$\omega$ by definition of $\eb$ given just before 
(\ref{eq-sec.hr.model.to2-junc.coeff}).

Here the following relation is important; 
$\E(x) = (1/2\sqrt{x}) \, e^{-x^2/2} \, \text{W}_{-1/4 , 1/4}(x^2)$, 
where $\text{W}_{\mu , \nu}(z)$ is the Whittaker function. 
Because the qualitative behavior of $\text{W}_{\mu , \nu}(z)$ is of 
dumping oscillation as a function of $z$, the normalized Hawking 
spectrum (\ref{eq-sec.hr.dn.norm-hr}) of present case is expected to 
be oscillatory with decreasing amplitude as a function of $\omega$. 
To estimate the high frequency amplitude, note that the asymptotic 
form of $\E(z)$ is given using that of $\text{W}_{\mu , \nu}(z)$ as 
$\E(z) \sim e^{-z^2}/2z$, as $|z| \to \infty$. Consequently we find 
$\DE(\Omega,\omega) \sim
   O(e^{-\omega^2/c\,\Omega} \, \sqrt{c\,\Omega}/\omega)$ as 
$\omega \to \infty$, then from (\ref{eq-sec.hr.model.to2-dev}), the 
high frequency amplitude of the Hawking spectrum can be roughly 
estimated to be 
$I_H(\eta_i,\eta_f;\omega) \sim O(e^{-2\omega^2}/\omega)$ as 
$\omega \to \infty$.

\subsubsection{Case: $T(\eta) - \Rt(\eta) \propto \eta^3$}
\label{sec.hr.model-to3}

The third case is that the scale factor satisfies, 
 \sikib
  T(\eta) - \Rt(\eta) = - c \, \eta^{\,3} + b \, \eta + p
 \quad , \quad
  c = \text{const} >0 \quad , \quad b, p = \text{const}.
 \sikie
Here note that even if there is a term proportional to $\eta^2$, 
it can be vanished by shifting the origin of $\eta$. The differential 
of this equation gives,
 \sikib
  \od{\,a(\eta)}{\eta} =
    \frac{\sqrt{1-k\,r_0^{\,\,2}}}{r_0} \, a(\eta)
  - \frac{-3\, c \, \eta^2 + b}{r_0} \,
     \left( 1- \frac{R_g}{a(\eta) \, r_0} \right) \, .
 \label{eq-sec.hr.model.to3-sfac}
 \sikie
From this equation we find that, if $a > 0$, the behavior of 
$a(\eta)$ is the exponential grow for $\eta>\sqrt{b/3c}$. That is, 
it can be said that this case also corresponds to the asymptotic form 
of the open FRW universe at a remote future, $a(t)\propto t$. As for 
the case of the previous subsection \ref{sec.hr.model-to2}, the 
difference from the thermal case comes from the detailed behavior of 
$a(t)$. For example, because $c>0$, the velocity of expansion 
$da/d\eta$ of the present case (\ref{eq-sec.hr.model.to3-sfac}) 
should be larger than that of the thermal case 
(\ref{eq-sec.hr.model.therm-sfac}) for $\eta > 0$. 

The spectrum (\ref{eq-sec.hr.dn.norm-hr}) of the present case is 
expected not to be the thermal one. To evaluate the normalized 
junction coefficient (\ref{eq-sec.hr.dn.norm-A1}), we should carry 
out the integral, 
$\int_{\eta_i}^{\eta_f} d\eta \,
   \exp[ic\, \Omega \eta^3 + i(\omega-b\, \Omega) \eta]$. But it is 
difficult at least for the author to evaluate this integral. 
Therefore in order to guess roughly the behavior of the Hawking 
radiation, we try using the non-normalized representation 
(\ref{eq-sec.hr.hr-hr}). The non-normalized junction coefficient 
(\ref{eq-sec.sf.mf.TRWX-coeff.A1}) is exactly calculated to be,
 \begin{equation*}
 \begin{split}
  A_{\Omega \omega}^{(1)} &= \,\,\,
   e^{-i \omega \rt_0 -i p \, \Omega} \,
    \sqrt{\left|\frac{\omega}{\Omega}\right|} \,
    \frac{\sqrt{|\omega - b\, \Omega|}}{[3(c\, \Omega)^{1/3}]^{3/2}}
 \\ &\qquad \times
    \left[\,
     J_{-1/3}\left[ 2 \left(
                    \frac{|\omega-b\, \Omega|}
                         {3(c\, \Omega)^{1/3}}\right)^{3/2} \right]
    - \sgn(\omega - b\, \Omega) \,
     J_{1/3}\left[ 2 \left(
                    \frac{|\omega-b\, \Omega|}
                         {3(c\, \Omega)^{1/3}}\right)^{3/2} \right]
    \, \right] \, ,
 \end{split}
 \end{equation*}
where the integral representation of the Bessel function is used,
 \sikibnon
  \int_{-\infty}^{\infty} dx \, e^{i (q x)^3 + i k x}
 =
  \frac{2\pi \sqrt{|k|}}{(3q)^{3/2}} \,
  \left[\,
    J_{-1/3}\left[ 2 \left(\frac{|k|}{3q}\right)^{3/2} \right]
  - \sgn(k) \,
    J_{1/3}\left[ 2 \left(\frac{|k|}{3q}\right)^{3/2} \right] \,
  \right]
 \quad \text{for $q>0$} \, .
 \sikienon
Then the deviation factor $D_H$ becomes, 
 \begin{equation*}
 \begin{split}
  D_H(\Omega,\omega)
 &= \frac{\omega}{27 c\, \Omega^2} \,
   \left[\,\,\, |\omega - b\, \Omega| \,
               \Delta_+(\Omega , \omega)^2
 + e^{2\pi\Omega/\sg} \, (\omega + b\, \Omega) \,
     \Delta_-(\Omega , \omega)^2 \right. \\
 & \qquad \qquad + \left.
   2 \, e^{\pi\Omega/\sg} \,
     \sqrt{|\omega^2-c^2\, \Omega^2|} \,
     \cos\left( 2 p \, \Omega \right) \,
     \Delta_+(\Omega , \omega) \, \Delta_-(\Omega , \omega)
   \,\,\, \right] \, ,
 \end{split}
 \end{equation*}
where
 \sikibnon
  \Delta_+(\Omega, \omega)
 &=& J_{-1/3}\left[ 2 \left(
                    \frac{|\omega - b\, \Omega|}
                         {3(c\, \Omega)^{1/3}}\right)^{3/2} \right]
    - \sgn(\omega - b\, \Omega) \,
     J_{1/3}\left[ 2 \left(
                    \frac{|\omega - b\, \Omega|}
                         {3(c\, \Omega)^{1/3}}\right)^{3/2} \right] \\
  \Delta_-(\Omega, \omega)
 &=& J_{-1/3}\left[ 2 \left(
                    \frac{\omega + b\, \Omega}
                         {3(c\, \Omega)^{1/3}}\right)^{3/2} \right]
   + J_{1/3}\left[ 2 \left(
                    \frac{\omega + b\, \Omega}
                         {3(c\, \Omega)^{1/3}}\right)^{3/2} \right]
 \, .
 \sikienon
This result denotes that the Hawking spectrum is oscillatory as a 
function of $\omega$ which is the frequency observed by a comoving 
observe at a remote future $\eta \to +\infty$. The amplitude can be 
roughly estimated to be 
$_W\dvac N_{F,\omega} \vac_W \sim \omega^{1/2}$ as 
$\omega \to \infty$, where the asymptotic form of the Bessel function 
is used, 
$J_{\nu}(z) \sim
  \sqrt{2/\pi z} \, \cos[z-(2\nu + 1)\pi/4]$ as $|z| \to \infty$.
Because this result is not normalized, the high frequency spectrum 
diverges. But the qualitative behavior that the spectrum is 
oscillatory may hold even for the normalized one.

\subsubsection{de-Sitter case: $a(\eta)=-c/\eta$}
\label{sec.hr.model-ds}

Here we turn our attention from the open universe to the other 
universe. Consider the de-Sitter case,
 \sikibnon
  a(\eta) = \frac{-c}{\eta}
 \quad , \quad \text{for $\eta<0$ and $c =$ const. $>0$}.
 \sikienon
The relation, $dt = a(\eta)d\eta$, gives $a(t)=\exp(t/c)$, that is, 
this is  the inflationary universe. 

From (\ref{eq-sec.scu.scu-time}) and the definition of $\Rt(\eta)$, 
we obtain the phase factor,
 \sikibnon
  T(\eta) - \Rt(\eta)
    = \frac{c \, r_0}{\eta}
     + \left( c \sqrt{1-k \,r_0^{\,\,2}} - R_g \right) \,
       \ln\left( \frac{1}{\eta} + \frac{R_g}{c \,r_0} \right)
     + p \, ,
 \sikienon
where $p$ is the integral constant due to 
(\ref{eq-sec.scu.scu-time}). Hereafter in this subsection, we assume 
that the junction surface $\js$ of the SCU is much larger than the 
black hole, $R_g \ll R(\eta) = -c \, r_0/\eta$. Then the 
approximation, $1/\eta + R_g/c\, r_0 \simeq 1/\eta$, becomes valid, 
and the normalized junction coefficient (\ref{eq-sec.hr.dn.norm-A1}) 
is expressed as
 \sikibnon
  A_{\Omega \, \omega}^{(1)} &\simeq&
    \frac{e^{-i\omega \rt_0 -ip\, \Omega}}{\eta_p} \,
    \sqrt{\left| \frac{\omega}{\Omega} \right|} \,
     \int_{\eta_i}^{\eta_f} d\eta \,
       e^{i(\omega \eta - c\,r_0 \,\Omega/\eta)} \,
       \eta^{i \Omega \left( c\,\sqrt{1-k\,r_0^{\,\,2}}
                            - R_g \right)} \\
 &=&
    \frac{e^{-i\omega \rt_0 -ip\, \Omega}}{\eta_p} \,
    \sqrt{\left| \frac{\omega}{\Omega} \right|} \,
    \left( \frac{c\,r_0 \, \Omega}{\omega}
    \right)^{-\nu} \,
     \int_{u_i}^{u_f} d u \, e^{z/2 \, (u-1/u)} \, u^{-1-\nu} \, ,
 \sikienon
where $z = i\, 2 \sqrt{c\,r_0 \, \Omega \, \omega}$, 
$\nu = -1 - i \Omega \,
              \left( c\,\sqrt{1-k\, r_0^{\,\,2}} - R_g \right)$, 
and the variable of integration is changed as 
$\eta \to u = \eta \, \sqrt{\omega/c r_0 \Omega}$. Here let us take 
the approximations, $\eta_f \to 0$ and $\eta_i \to -\infty$, and 
further consider the modification, 
$\omega \to \lim_{\epsilon \to 0} \omega \, e^{-i \epsilon}$. Then 
the integral path can be analytically connected to the complex 
$u$-plane and deformed to $C$ shown at the figure \ref{fig-path}, 
which gives 
 \sikib
  A_{\Omega , \omega}^{(1)} \simeq
    \frac{-i\pi e^{i\omega \rt_0 -ip\, \Omega}}{\eta_p} \,
    \left( \frac{c\, r_0 \, \Omega}{\omega}
    \right)^{-\nu/2} \, 
    \sqrt{\left| \frac{\omega}{\Omega} \right|} \,\,\,
    \Han_{\nu}^{(1)}
      \left( i \, 2 \sqrt{c\, r_0 \, \Omega \, \omega} \right) \, ,
 \label{eq-sec.hr.model.ds-jc}
 \sikie
where the integral representation of the Hankel function is used,
$i \pi \Han_{\nu}^{(1)}(z) = - \int_{C} d u \,
    e^{z/2 \, (u-1/u)} \, u^{-1 - \nu}$ for $\Re(z)>0$. 
This is valid only for $\omega > 0$. 
But we can obtain the same result (\ref{eq-sec.hr.model.ds-jc}) for 
$\omega < 0$ by routine calculations with taking the branch cut of 
$\ln u$ in the upper half $u$-plane ($-1 = e^{-i \pi}$). Therefore 
the deviation factor is given as
 \sikibnon
  D_H(\Omega , \omega)
 &\simeq& \frac{\pi^2 c\, r_0}{\eta_p^{\,\,2}} \,
   \left[ \left|
            \Han_{\nu}^{(1)}
            \left( i \, 2 \sqrt{c\,r_0 \, \Omega \, \omega} \right)
          \right|^2
       + e^{2\pi \Omega/\kappa} \,
          \left|
            \Han_{\nu}^{(1)}
            \left( 2 \sqrt{c\,r_0 \, \Omega \, \omega} \right)
          \right|^2 \right. \\
  &\quad& \left.
    + 2\, e^{\pi \Omega/\kappa} \,
      \Re\left( e^{i\pi/2-i 2 p\, \Omega}
                \left( \frac{c\,r_0 \, \Omega}{\omega}
                \right)^{-1-\nu} \,
           \Han_{\nu}^{(1)}
            \left( i \, 2 \sqrt{c\,r_0 \, \Omega \, \omega} \right) \,
           \Han_{\nu}^{(1)}
            \left( 2 \sqrt{c\,r_0 \, \Omega \, \omega} \right)
         \right) \right] \, .
 \sikienon
This is essentially the dumping oscillation about $\omega$. 

The amplitude of the high frequency spectrum is estimated using the 
asymptotic form of the Hankel function, 
$\Han_{\nu}^{(1)}(z) \sim \sqrt{2/\pi z} \, e^{i(z-(2\nu+1)\pi/4)}$, 
to be, 
$\Han_{\nu}^{(1)}\left( i 2 \sqrt{c\,r_0 \, \Omega \, \omega} \right)
  \sim O\left( (\Omega \omega)^{-1/4}
        \times e^{-\sqrt{\Omega \omega}-\Omega} \right)$. Then the 
dominant term of $D_H$ is the second term, and we find that 
$I_H(-\infty,0 ; \omega) \sim O(1/\sqrt{\omega})$ as 
$\omega \to \infty$.

\subsubsection{Case: close to the thermal spectrum}
\label{sec.hr.model-close}

Finally let us examine the case which is close to the thermal case. 
According to the subsection \ref{sec.hr.model-therm}, the Hawking 
spectrum (\ref{eq-sec.hr.dn.norm-hr}) should be reduced to the 
thermal one by a manipulation which makes the scale factor come 
to satisfy (\ref{eq-sec.hr.model.therm-sfac}). For example, the slow 
cosmological expansion limit, $\dot{a}/a \to 0$, would reproduce the 
thermal spectrum. Therefore we consider the case, 
 \sikibnon
  T(\eta) - \Rt(\eta) = c \, \eta^l + b\, \eta + p
  \quad \text{as $c \to 0$},
 \sikienon
where $l$ is an arbitrary natural number, and it is assumed that 
the scale factor is analytic as a function of $\eta$ and consequently 
the phase factor $T-\Rt$ is also analytic. This manipulation, 
$c \to 0$, should include the limit $\dot{a}/a \to 0$. 

As $c \to 0$, the junction coefficient (\ref{eq-sec.hr.dn.norm-A1}) 
is calculated as,
 \sikibnon
   A_{\Omega \, \omega}^{(1)} &\simeq&
    \frac{e^{-i\omega \rt_0 -ip\, \Omega}}{\eta_p} \,
    \sqrt{\left| \frac{\omega}{\Omega} \right|} \,
     \int_{\eta_i}^{\eta_f} d\eta \,
       \left( 1 - i c \, \Omega \, \eta^l \right) \,
       e^{i(\omega - b \,\Omega) \eta} \\
  &=&
   e^{-i\omega \rt_0 -ip\, \Omega} \,
   \sqrt{\left| \frac{\omega}{\Omega} \right|} \,
    \left[ \delta_{\omega , b \Omega}
         - \frac{i c \, \Omega}{\eta_p} \,
             \int_{\eta_i}^{\eta_f} d\eta \, \eta^l \,
              e^{i(\omega - b \,\Omega) \eta} \right] \, .
 \sikienon
Here by treating the Kronecker $\delta_{\omega , b \Omega}$ as a 
distribution and referring to the property of the delta function, 
$d\delta(x)/dx = -\delta(x) d/dx$, the integral in the second term 
becomes,
 \sikibnon
   \int_{\eta_i}^{\eta_f} d\eta \eta^l \,
    e^{i(\omega - b \,\Omega) \eta}
 = \eta_p \left( \frac{1}{i b} \right)^l \,
    \delta_{\omega , b \Omega} \, \left( \od{}{\Omega} \right)^l \, ,
 \sikienon
which gives
 \sikibnon
  A_{\Omega \, \omega}^{(1)} \simeq
    e^{-i\omega \rt_0 -ip\, \Omega} \,
    \sqrt{\left| \frac{\omega}{\Omega} \right|} \,
      \delta_{\omega , b \Omega} \,
      \left[ 1 + (-i)^{l+1} \, \frac{c \, \Omega}{b^l} \,
                  \left( \od{}{\Omega} \right)^l \right] \, ,
 \sikienon
where $\omega = 2\pi n /\eta_p$, $\Omega = 2\pi N/T_p$. This means 
$A_{\Omega \, -\omega}^{(1)} = 0$ for $\omega>0$ and $\Omega>0$, 
therefore we obtain 
 \sikibnon
  && D_H(\Omega , \omega) =
      \left| A_{\Omega \, \omega}^{(1)} \right|^2 \quad \to \quad
       \frac{\omega}{\Omega} \, \delta_{\omega , b \Omega}^{\,\,\, 2}
     \quad \text{as $c \to 0$} \\
  &\Longrightarrow&
   I_H(\eta_i,\eta_f ; \omega) \quad \to \quad
    \frac{b}{e^{2\pi\omega/b\sg}-1} \quad \text{as $c \to 0$}.
 \sikienon
In the limit $c \to 0$, the second and third terms of the deviation 
factor $D_H$ do not contribute to the Hawking spectrum 
$I_H(\eta_i,\eta_f ; \omega)$, and the first term of $D_H$ is reduced 
to that of the thermal case. However in the non-linear case, $c \ne$ 
so small, the second and third terms of $D_H$ come to contribute to 
$I_H(\eta_i,\eta_f ; \omega)$. That is, we can expect naively that 
the intensity of the non-thermal Hawking radiation is stronger than 
that of the thermal one due to the exponential factor 
$e^{\pi \Omega/\sg}$ in $D_H$, where the mass of the black hole is 
the same for both of the cases.


\section{Summary and discussion}
\label{sec-sd}

\subsection{Summary}
\label{sec.sd-sum}

{\it How does a dynamical boundary condition like an expanding 
universe change the well-known properties of asymptotically flat 
black hole spacetimes?} This paper is the first trial to treat this 
question, and deals with the effect of cosmological expansion on the 
Hawking radiation. The swiss cheese universe (SCU) has been adopted 
as a concrete model of background spacetime. With using the SCU we 
could avoid the issue how to define the black hole in an expanding 
universe, and the mass of the black hole could be defined by the 
Kumar mass computed on the spatial section of the junction surface 
$\js$ of the SCU. Further for simplicity, the SCU has been assumed to 
be two dimensional in order to neglect the curvature scattering of a 
massless scalar field of minimal coupling. It has also been assumed 
that the SCU was eternal, that is, no big bang and big crunch 
singularities would appear and the scale factor $a(\eta)$ could be 
defined in the infinite interval of the conformal time of the 
FRW-side, $-\infty < \eta < \infty$. 

We have introduced a massless scalar field $\SFD$ of minimal 
coupling, and ignored the back reaction to the background SCU. The 
junction condition of $\SFD$ has required no potential confined on 
$\js$. Therefore $\SFD$ was not reflected at $\js$. Consequently, 
since we have neglected the curvature scattering, a mode function 
which is a monochromatic out-going one in the BH-side has been 
connected at $\js$ to a mode function in the FRW-side which is of a 
superposition of the out-going modes without including any in-going 
one, and so has been the in-going modes. 

The scalar field has been quantized with the WX-mode, the TR-mode and 
the Cfl-mode. Because the event horizon bifurcated in the eternal 
SCU, we could make use of the discussion given for the asymptotically 
flat eternal Schwarzschild spacetime, and obtained the Bogoljubov 
transformation from the WX-mode to the TR-mode \cite{ref-unruh}. For 
finding the Bogoljubov transformation from the TR-mode to the 
Cfl-mode, it has been essential that the junction coefficients of the 
mode functions determined at $\js$ could be re-interpreted as the 
Bogoljubov coefficients. Finally by the successive Bogoljubov 
transformation from the WX-mode to the Cfl-mode via the TR-mode, we 
have obtained the Hawking radiation (\ref{eq-sec.hr.hr-hr}) as the 
particle of the Cfl-mode created on the vacuum state of the WX-mode. 
This Hawking radiation was measured by a comoving observer at a 
remote future $\eta \to +\infty$. 

However, because the back reaction was ignored, the Hawking spectrum 
(\ref{eq-sec.hr.hr-hr}) has diverged as mentioned at the beginning of 
the section \ref{sec.hr-dn}. This divergence arose from the infinite 
time interval of the observation. Therefore we have introduced the 
finite time interval normalization which restricted the observation 
time, $\eta_i < \eta < \eta_f$, and obtained the normalized Hawking 
spectrum (\ref{eq-sec.hr.dn.norm-hr}).

\subsection{Discussions}
\label{sec.sd-dis}

It is easily expected that the Bogoljubov transformation from the 
TR-mode to the Cfl-mode causes the difference of the resultant 
spectrum from the thermal one. Note that this Bogoljubov 
transformation does not include the cosmological particle creation, 
since the massless scalar field of minimal coupling in two 
dimensional spacetime is not scattered by the background curvature. 
Therefore the deviation factor $D_H$ of the resultant spectrum 
(\ref{eq-sec.hr.hr-dev}) includes only the effect of the relative 
motion of the comoving observer to the TR-system. One may think that 
the resultant spectrum would be just a thermal one which would 
receive a red shift to a lower temperature due to the comoving 
motion. However it is obvious from (\ref{eq-sec.hr.dn.norm-hr}) that 
the resultant Hawking spectrum is totally different from a thermal 
one, except the case of subsection \ref{sec.hr.model-therm}. The 
reason why the spectrum is not generally a thermal one, is in the 
junction condition of $\SFD$ at $\js$. If a mode function can be a 
monochromatic in both of the BH-side and the FRW-side, we can 
understand the junction condition as a simple Doppler effect. However 
as mentioned in the previous subsection, a monochromatic out-going 
mode in the BH-side goes through $\js$ into the FRW-side, then it 
turns to a superposition of out-going modes in the FRW-side. 
Therefore a monochromatic out-going mode received by a comoving 
observer in the FRW-side, should be a superposition of out-going 
modes in the BH-side when it would be traced back into the BH-side. 
This denotes that various frequencies $\Omega$ of the TR-mode are 
included in each monochromatic mode of frequency $\omega$ in the 
FRW-side. That is, so many Doppler effects take place at the same 
time in one Cfl-mode, which is represented by $D_H$. We interpret 
such the superposition of Doppler effects as the pure effect of the 
acceleration due to the cosmological expansion. 

According to the subsection \ref{sec.hr.model-therm}, it seems that 
the thermal Hawking spectrum is of a special case where the second 
and third terms in the representation (\ref{eq-sec.hr.hr-dev}) of 
$D_H$ vanish, and that the Hawking spectrum 
(\ref{eq-sec.hr.dn.norm-hr}) is generally different from the thermal 
one. However as examined in the subsection \ref{sec.hr.model-close}, 
a non-thermal spectrum is reduced to the thermal one by an 
appropriate manipulation, for example, by the slow cosmological 
expansion limit, $\dot{a}/a \to 0$. About the non-thermal Hawking 
spectrum, the subsections \ref{sec.hr.model-to2}, 
\ref{sec.hr.model-to3} and \ref{sec.hr.model-ds} give the conclusion 
that the qualitative behavior of the non-thermal spectrum is the 
dumping oscillation as a function of the frequency $\omega$ measured 
by a comoving observer. Further as discussed at the end of the 
subsection \ref{sec.hr.model-close}, the intensity of the non-thermal 
Hawking radiation is stronger than that of the thermal one. Then we 
can describe a picture from the viewpoint of the black hole 
thermodynamics that {\it a black hole with an asymptotically flat 
boundary condition stays in a lowest energy thermal equilibrium 
state. When a black hole is put into a dynamical boundary condition, 
it is excited to a non-equilibrium state, and emits its mass energy 
with stronger intensity than the thermal one.} However, about our 
general aim given at the beginning of section \ref{sec-intro}, we 
could not search for a concrete example of the non-gravitational and 
non-equilibrium system which corresponds to a black hole in an 
expanding universe. 

So far we have assumed two dimensional SCU. In extending to four 
dimension, there arise three problems due to the curvature 
scattering: (i) the initial vacuum state, (ii) the junction of $\SFD$ 
at $\js$ and (iii) the cosmological particle creation. For the first 
issue, as mentioned in the subsection \ref{sec.scu.str-second}, we 
should modify the vacuum state on which the particle creation is 
estimated. The second issue means that, even if $F=0$ at 
(\ref{eq-app.jc-jcsf.diff}), a reflection of $\SFD$ due to the 
curvature scattering takes place at $\js$, that is, a monochromatic 
out-going mode in the BH-side should be a superposition of the 
out-going and in-going modes in the FRW-side. Both of these issues 
(i) and (ii) do not cause any change for our strategy explained in the 
section \ref{sec.scu-str}. However about the third issue (iii), we 
should add another step to our strategy. The new step is an extra 
Bogoljubov transformation from the Cfl-mode of initial time to that 
of the final time, which raises the cosmological particle creation. 
But such a Bogoljubov transformation will take a complicated form, 
and we can hardly estimate its form since it is difficult to solve 
the mode functions with considering the curvature scattering. However 
if one wants to research the effects of only the cosmological 
particle creation, it is the easiest model to consider the massive 
scalar field on two dimensional SCU \cite{ref-bd}.

Finally let us discuss about the assumption of the ``eternal'' SCU. 
As mentioned at the end of section \ref{sec.scu-scu}, though the 
evolution of the scale factor is mathematically given by the equation 
(\ref{eq-app.scu-sfac}), we dared to assume that the scale factor was 
a positive definite function $a(\eta)>0$ for 
$-\infty < \eta < \infty$. With this assumption we could obtain the 
form of the junction coefficients $A^{(1,2)}$ in the simple 
representation of (\ref{eq-sec.sf.mf.TRWX-coeff.out}). Then what 
should we modify if the equation (\ref{eq-app.scu-sfac}) is taken 
into account? Because of the initial singularity $a(t=0)=0$, we 
should solve the junction conditions of the scalar field $\SFD$ for 
the interval $\eta_0 < \eta < \infty$, where $\eta_0$ is given by 
$a(\eta_0)\, r_0 = R_g$. This modifies the completeness 
(\ref{eq-sec.sf.mf.prep-Cfl.comp}) to 
 \sikibnon
  \int_{\eta_0}^{\infty} d\eta \,
         f_{F,\omega}^{\ast}(\eta) \, f_{F,\omega'}(\eta)
    = \pi \, \delta(\omega-\omega')
    + i e^{i(\omega'-\omega)\eta_0} \,\pv \, 
       \frac{1}{\omega-\omega'} \, ,
 \sikienon
where $\pv$ is the Cauchy's principle value. The second term may 
raise some complicated term in the equation 
(\ref{eq-sec.sf.mf.TRWX-coeff.out}), which would make the following 
computations intricate. However if we can assume that 
 \sikib
  \int_{-\infty}^{\infty} d\omega' \, \pv \,
    \frac{A_{\Omega \omega'}^{(1,2)}}{\omega-\omega'} \,
      \phi_{F,\omega'}^{(1,2)} = 0 \, ,
 \label{eq-sec.sd-assume}
 \sikie
then we can obtain the junction coefficients of the form 
(\ref{eq-sec.sf.mf.TRWX-coeff.out}) with modifying the interval 
of integration of $I_A$ and $J_A$ from $-\infty < \eta < \infty$ to 
$\eta_0 < \eta < \infty$. Further even if this modified junction 
coefficient makes the Hawking spectrum diverge, we can expect well 
that the same normalization method discussed in section 
\ref{sec.hr-dn} holds. Here, by considering the four dimensional 
case, we can find the validity of the assumption 
(\ref{eq-sec.sd-assume}). In four dimensional case, we can also 
separate the variables on the FRW-side, 
$\SFD = f_F(\eta) \, h_{F}(\rt) \, Y_{lm}(\theta,\varphi)$, where 
$Y_{lm}$ is the spherical harmonics. It can be easily found that the 
portion $f_F(\eta)$ satisfies a Sturm-Liouville type ordinary 
differential equation. This implies that there is a complete 
orthogonal set $\{ f_{F, \omega} \}$ satisfying 
(\ref{eq-sec.sf.mf.prep-Cfl.comp}) with appropriate range and 
measure about $\eta$ and $\omega$. Therefore it is not so bad to 
assume the background spacetime is the eternal SCU in order to 
extract the essence of the Hawking radiation in an expanding 
universe.


\section*{Acknowledgements}

I would like to thank to M.Sakagami and A.Ohashi for their helpful 
comments and discussions, and to J.Soda for his useful comments. 
Further I wish to express my gratitude to W.Israel, J.Ho and 
S.Mukohyama for their useful discussions during my temporary stay at 
the University of Victoria in Jun.2001. This work is supported by 
JSPS Research Fellowships for Young Scientists.

\appendix

\section{Four dimensional swiss cheese universe}
\label{app-scu}

The swiss cheese universe (SCU) \cite{ref-scu} is the spacetime 
including a spherically symmetric black hole in an expanding 
universe, and obtained by connecting a Schwarzschild spacetime with a 
dust-dominated Friedmann-Robertson-Walker (FRW) spacetime at a given 
spherically symmetric timelike hypersurface, $\js$, by the Israel 
junction condition with no energy density confined on the junction 
surface \cite{ref-israel}. In this appendix, we show a sketch of 
constructing the SCU in four dimension. Hereafter as the terminology, 
let the word ``BH-side'' denote the spacetime region inside $\js$ 
where the metric is given by (\ref{eq-app.scu-BH}), and ``FRW-side'' 
for the region outside $\js$ where the metric is 
(\ref{eq-app.scu-FRW}).

The metric of a Schwarzschild black hole is
 \sikib
  ds_{BH}^2 =
   - \left( 1-\frac{R_g}{R} \right) dT^2
   + \left( 1-\frac{R_g}{R} \right)^{-1} dR^2
   + R^2 \left( d\theta^2 + \sin^2\theta d\varphi^2 \right) \, ,
 \label{eq-app.scu-BH}
 \sikie
where $x_{BH}^{\mu} = (T,R,\theta,\varphi)$ is the Schwarzschild 
coordinate and $R_g = 2GM$, $M$ is the mass of the black hole and $G$ 
is the gravitational constant. The metric of a FRW universe is 
 \sikib
  ds_F^2 =
   - dt^2
   + a(t)^2 \left[ \frac{dr^2}{1-k r^2}
      + r^2 \left( d\theta^2
           + \sin^2\theta d\varphi^2 \right) \right] \, ,
 \label{eq-app.scu-FRW}
 \sikie
where $x_F^{\mu} = (t,r,\theta,\varphi)$ is the comoving coordinate, 
$a(t)$ is the scale factor and $k=\pm1$, $0$ is the spatial 
curvature. We can make the angular coordinates $\theta$ and $\varphi$ 
to be common to both of these coordinates due to the spherical 
symmetry. Further we assume that the Friedmann equation, that is the 
evolution of $a(t)$, is of dust-dominated one,
 \sikib
  \left( \frac{\dot{a}}{a} \right)^2 + \frac{k}{a^2}
     = \frac{8\pi G}{3} \, \frac{\rho_0}{a^3} \, ,
 \label{eq-app.scu-sf}
 \sikie
where $\rho_0$ ($=$ const.) is of free parameter at this moment, but 
will be fixed consistently in considering the junction condition 
later. 

In constructing the SCU we choose $\js$ to be spherically symmetric 
and timelike. Therefore the angular coordinate $\theta$ and $\varphi$ 
can be chosen as the spatial coordinates on $\js$, and it is possible 
to use $t$ as the temporal coordinate on $\js$. That is, the 
coordinate we use in $\js$ is expressed as 
$u^{i} = (t,\theta,\varphi)$.
\footnote{It is also possible to use $T$ as the temporal coordinate 
on $\js$. The reason we chose $t$ is just the later convenience.}
The radial coordinates of $\js$ measured from BH-side and FRW-side 
are generally given as a function of $t$, like $R(t)$ and $r(t)$ 
respectively. However especially about $r(t)$, we can specify it to 
be constant, $r=r_0=$ constant. To understand this feature of radial 
coordinate of $\js$, it is important to notice that the mass of the 
black hole $M$ is assumed to be constant. If $r(t) \neq$ constant, 
there should be an energy flow of dust-matter through $\js$, since 
the dust-matter rests on the comoving coordinate. Consequently, 
because it is also assumed that no energy density is confined on 
$\js$, the source of the energy flow should be the mass of black 
hole, that is, $M$ should have time dependence. 
\footnote{For more correct discussion, consider the Komar integral 
evaluated on an arbitrary closed two dimensional spatial surface 
including the spatial section of the event horizon. Then the 
resultant Komar mass is exactly $M$ for the metric 
(\ref{eq-app.scu-BH}) as calculated at the end of this appendix 
\ref{app-scu}. This means that the energy flow through $\js$ affects 
directly the mass of the black hole, if $\js$ is located outside the 
event horizon.}
This contradicts $M =$ constant. Hence $r(t) = r_0 =$ constant. On 
the other hand, another function $R(t)$ cannot be specified at this 
moment, but will be determined by the junction condition. 

The Israel junction condition at the junction surface $\js$ with no 
energy density confined on $\js$, is given by \cite{ref-israel}
 \sikib
   {\bf h}_{BH}|_{\js} &=& {\bf h}_F|_{\js}
     \label{eq-app.scu-jc.metric} \\
   {\bf K}_{BH}|_{\js} &=& {\bf K}_F|_{\js} \, ,
     \label{eq-app.scu-jc.curv}
 \sikie
where ${\bf h}_{BH}|_{\js}$ and ${\bf K}_{BH}|_{\js}$ are 
respectively the induced metric and the extrinsic curvature of $\js$ 
measured from the BH-side, and similarly to ${\bf h}_F|_{\js}$ and 
${\bf K}_F|_{\js}$ with respect to the FRW-side. The components of 
${\bf h}$ and ${\bf K}$ on $\js$ are respectively, 
$h_{ij} = g_{\mu \nu}\, x^{\mu}_{,i} \, x^{\nu}_{,j}$, and 
$K_{ij} = n_{\mu ; \nu}\, x^{\mu}_{,i} \, x^{\nu}_{,j}$, where 
$x^{\mu}_{,i} = \partial x^{\mu}/\partial u^i \, $, $g_{\mu \nu}$ is 
the metric of four dimension, $n^{\mu}$ is the unit vector normal to 
$\js$. In the FRW-side, the unit normal to $\js$ is obtained from the 
expression of $\js$, $S_F \equiv r-r_0 = 0$, 
 \sikib
   n_F^{\mu}
 = \frac{S_F^{\,\,\, ,\mu}}
        {\sqrt{|S_{F,\alpha} S_F^{\,\,\, ,\alpha}|}}
 = \left( 0 \, , \,
      \frac{1}{a(t)}\sqrt{1-k r_0^{\,2}} \, , \,
          0 \, , \, 0 \right)
 \label{eq-app.scu-normal.FRW} \, .
 \sikie
In the BH-side, $S_{BH} \equiv R - R(t) = 0$ gives $\js$. Using the 
the junction condition (\ref{eq-app.scu-jc.metric}), we obtain the 
unit normal to 
$\js$,
 \sikib
   n_{BH}^{\mu}
 = \frac{S_{BH}^{\,\,\, ,\mu}}
        {\sqrt{|S_{BH,\alpha} S_{BH}^{\,\,\, ,\alpha}|}}
 = \left( \left[ 1-\frac{R_g}{R(t)} \right]^{-1} \dot{R}(t) \, , \, 
          \left[ 1-\frac{R_g}{R(t)} \right] \dot{T}(t) \, , \, 
          0 \, , \, 0 \right) \, .
 \label{eq-app.scu-normal.BH}
 \sikie
The results of the junction conditions (\ref{eq-app.scu-jc.metric}) 
and (\ref{eq-app.scu-jc.curv}), are summarized into three independent 
equations. One of them specifies the location of $\js$ in the 
BH-side, $R(t) = a(t) \, r_0$ in BH-side. The other two equations 
give the form of the function $T(t)$ and the evolution of the scale 
factor on $\js$,
 \begin{gather}
  \od{T(t)}{t}
     = \frac{\sqrt{1-k r_0^{\,\, 2} }}{1-R_g/R(t)} \, ,
      \label{eq-app.scu-T} \\
  \left( \frac{\dot{a}}{a} \right)^2 + \frac{k}{a^2}
     = \frac{Rg}{r_0^{\,\, 3}} \, \frac{1}{a^3}
      \label{eq-app.scu-sfac} \, ,
 \end{gather}
Comparing the result (\ref{eq-app.scu-sfac}) with the assumption 
(\ref{eq-app.scu-sf}), the consistent choice of the constant $\rho_0$ 
is
 \sikib
  \rho_0 = \frac{M}{(4/3)\pi r_0^{\, 3}} \, .
 \label{eq-app.scu-dust}
 \sikie

\begin{figure}[t]
  \cenb
   \includegraphics[height=30mm]{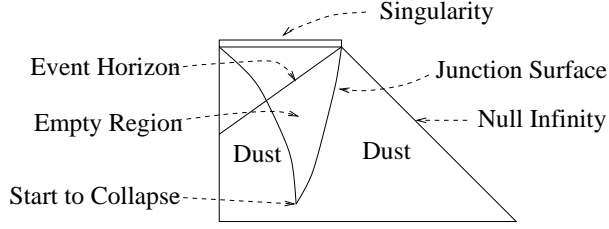}
  \cene
  \caption{{\small Conformal diagram of the SCU interpreted as the 
collapse of the dust-matter in flat or open FRW universe.}}
 \label{fig-scu.collapse}
 \end{figure}

To draw a picture of the SCU, assume $R(t_c)>R_g$ holds at an 
arbitrary time $t_c$ for the case that the FRW-side is open $k=-1$ or 
flat $k=0$. Then, because $a(t)$ increases monotonically in these 
cases, the relation (\ref{eq-app.scu-dust}) provides us with the 
following picture of the SCU: in a dust-dominated FRW universe 
including no black hole, the dust-matter contained in a spherically 
symmetric region of comoving radius $r_0$ starts to collapse, and 
finally a spherically symmetric black hole of mass 
$M = (4/3)\pi r_0^{\, 3} \rho_0$ is produced in an expanding 
universe. The conformal diagram of this interpretation is shown at 
the figure \ref{fig-scu.collapse}, which makes it clear that a 
black hole can be defined in the SCU with open and flat FRW-side. On 
the other hand, it is impossible to define a black hole for the case 
of closed FRW-side. Because $a(t)$ decreases to zero in a finite 
future, no causal curve cannot avoid encountering the big crunch 
singularity, hence every spacetime point is included in the causal 
past of the singularity and we cannot define a black hole. 

Finally as the definition of the mass of the black hole in the SCU, 
we adopt the Komar mass calculated on the junction surface $\js$. The 
general form of the Komar mass $Q_K$ \cite{ref-townsend} is
 \sikibnon
   Q_K = \frac{1}{8\pi G} \oint_{\partial A}
    dS_{\mu \nu} \xi^{\mu ; \nu} \, ,
 \sikienon
where $A$ is a spacelike region, $\partial A$ is the boundary of $A$, 
${\bf \xi}$ is the timelike Killing vector in $A$ and $dS_{\mu \nu}$ 
is the surface area form on $\partial A$. Using the Einstein equation 
and the Killing equation, it can be proved that $Q_K$ is invariant 
under any deformation of $A$ if ${\bf \xi}$ lasts to exist on the 
deformed region. For the SCU, $\partial A$ is the spatial section of 
$\js$ at $T=$ constant, and we can let ${\bf \xi} = \partial_T$, then 
the Komar mass is obtained to be $Q_K = M$. Here note that an 
important issue is left: the normalization of our Killing vector 
${\bf \xi}=\partial_T$ is not given at the junction surface nor at 
the spatial infinity of the FRW-side, but given at the asymptotically 
flat region of the full Schwarzschild spacetime which is not present 
in the SCU. Therefore we should say that, with accepting the vector 
$\partial_T$ as a timelike Killing vector in the BH-side, the $M$ 
appeared in (\ref{eq-app.scu-BH}) can be understood as a mass of the 
black hole in the SCU which is invariant under any deformation of the 
spatial hypersurface in the SCU. Further as a by-product of this 
normalization, the surface gravity $\sg$ of the event horizon is 
evaluated to be $\sg = 1/2R_g$.

\section{Junction condition for scalar field}
\label{app-jc}

The junction condition for connecting a scalar field $\sfd$ at a 
given hypersurface $\js$ is discussed in this appendix. As a 
terminology, we call the spacetime regions separated by $\js$ 
A-region and B-region, and denote a quantity $Q$ evaluated in 
A-region and B-region as $Q_A$ and $Q_B$ respectively. We consider 
the four dimensional background spacetime in the Gaussian normal 
coordinate with respect to a hypersurface $\js$, and the scalar field 
$\sfd$ with an arbitrary potential $V(\sfd)$. The metric is expressed 
as
 \sikibnon
  ds^2 = \frac{1}{n^2} \, dx^{n\,2} + g_{ij} \, dx^i dx^j \, ,
 \sikienon
where $x^i$ is the coordinate intrinsic to the hypersurface $\js$, 
$x^n$ is the coordinate vertical to $\js$, and $n^{\mu}=g^{n \mu}$ is 
the unit normal vector to $\js$, that is, $n^2=1$ if $\js$ is 
timelike or $n^2=-1$ if $\js$ is spacelike. Hereafter we set the 
surface $\js$ is placed at $x^n=0$. The Klein-Gordon equation is
 \sikibnon
  \square \sfd - \od{V(\sfd)}{\sfd}= 0 \, .
 \sikienon
By integrating this equation along $x^n$ direction through $\js$, we 
find 
 \sikibnon
   \int_{-\epsilon}^{\epsilon} dx^n \, \square \sfd
     = \int_{-\epsilon}^{\epsilon} dx^n \, \od{V}{\sfd} \, .
 \sikienon
On the other hand by definition of the d'Alembertian, we obtain 
 \sikibnon
   \int_{-\epsilon}^{\epsilon} dx^n \, \square \sfd
     = \int_{-\epsilon}^{\epsilon} dx^n \,
            \partial_n \partial^n \sfd 
       + \int_{-\epsilon}^{\epsilon} dx^n 
             \left( \Gamma^n_{n \mu} \partial^{\mu} \sfd
                  + \nabla_i \nabla^i \sfd \right)
     \,\,\, \longrightarrow \,\,\,
     \partial^n \sfd_A - \partial^n \sfd_B
         \quad \text{as $\epsilon \to 0$} \, ,
 \sikienon
where $\partial^n = n^{\mu}\partial_{\mu} \sfd$, and it is assumed 
that the A-region is the region of $x^n>0$ and B-region is of 
$x^n<0$. Here assume that there is a potential confined on the 
surface $\js$ such as 
 \sikibnon
   \od{V(\sfd)}{\sfd} = F \delta(x^n) \, ,
 \sikienon
where $F$ denotes a ``surface potential force'' derived from the 
potential confined on $\js$. Further it is natural to require the 
continuity of $\sfd$ on $\js$. Hence finally we obtain the junction 
condition of a scalar field on a given surface $\js$ to be
 \begin{gather}
  \sfd_A |_{\js} = \sfd_B |_{\js}
 \label{eq-app.jc-jcsf.sf} \\
  n_A^{\mu}\partial_{\mu} \sfd_A |_{\js}
   - n_B^{\mu}\partial_{\mu} \sfd_B |_{\js} = F \, ,
 \label{eq-app.jc-jcsf.diff}
 \end{gather}
If no potential is confined on the surface $\js$, the surface force 
vanishes $F=0$. This junction condition of a scalar field is very 
similar to the Israel junction condition for connecting geometries of 
spacetimes \cite{ref-israel} in the sense that a kind of source of 
the field under consideration is required in order to satisfy the 
field equation even on the junction surface.

\section{Second term of equation (\ref{eq-sec.sf.mf.TRWX-osc.out})}
\label{app-terms}

In this appendix we discuss how the second term of 
(\ref{eq-sec.sf.mf.TRWX-osc.out}) vanishes and what it means. To 
begin with, we need to know the behavior of the phase 
$-i \, \Omega [T(\eta) - \Rt(\eta)]$. Because the background is the 
eternal SCU, the junction surface $\js$ reaches the past temporal 
infinity as $\eta \to -\infty$ and the future temporal infinity as 
$\eta \to +\infty$. Then the equations 
(\ref{eq-sec.scu.scu-coordtrans.BH}) give 
$T(\eta) - \Rt(\eta) \rightarrow \pm \infty$ as 
$\eta \to \pm \infty$. 

Note that the mode functions $\psi_{BH,\Omega}^{(\pm)}$ should be the 
basis of the general solution of the Klein-Gordon equation,
 \sikibnon
  \SFD = \int_{-\infty}^{\infty} d\Omega \, \left[
          E_{\Omega}^{(+)} \, \psi_{BH,\Omega}^{(+)}
        + E_{\Omega}^{(-)} \, \psi_{BH,\Omega}^{(-)} \right] \, ,
 \sikienon
where the coefficients $E^{(\pm)}$ depend only on $\Omega$. This 
means that we should consider the property of the modes 
$\psi_{BH,\Omega}^{(\pm)}$ under the integration with a function of 
$\Omega$. Let $H(\Omega)$ be a test function, we can calculate as
\footnote{Mathematically this is the Riemann-Lebesgue Integral 
Theorem.}
 \begin{multline*}
  \int_{-\infty}^{\infty} d\Omega \,
    \frac{H(\Omega)}{\sqrt{|\Omega|}} \, 
      e^{-i\, \Omega[T(\eta) - \Rt(\eta)]} \\
  =  \left.
    \frac{i}{T(\eta) - \Rt(\eta)} \,
      \frac{H(\Omega)}{\sqrt{|\Omega|}} \, 
        e^{-i\, \Omega[T(\eta) - \Rt(\eta)]}
     \right|_{\Omega=-\infty}^{\Omega=+\infty}
  - \int_{-\infty}^{\infty} d\Omega \,
      \frac{i}{T(\eta) - \Rt(\eta)} \,
        \left( \od{}{\Omega} \frac{H(\Omega)}{\sqrt{|\Omega|}} 
        \right) \, e^{-i\, \Omega[T(\eta) - \Rt(\eta)]} \, ,
 \end{multline*}
then because $T(\eta) - \Rt(\eta) \rightarrow \pm \infty$ as 
$\eta \to \pm \infty$, 
 \sikibnon
  \int_{-\infty}^{\infty} d\Omega \,
    \frac{H(\Omega)}{\sqrt{|\Omega|}} \, 
      e^{-i\, \Omega[T(\eta) - \Rt(\eta)]} = 0
 \quad \text{as $\eta \to \pm \infty$} \, .
 \sikienon
Hence the second term of (\ref{eq-sec.sf.mf.TRWX-osc.out}) vanishes
\footnote{In an exact word, this is the weak convergence.}
whenever the mode functions $\psi_{BH,\Omega}^{(\pm)}$ are considered 
under the integration with respect to $\Omega$, 
 \sikib
  \frac{\sgn(\eb) \, \pi}{(4\pi)^{3/2} \sqrt{|\Omega|}} \, 
   \left. e^{-i\, \Omega[T(\eb) - \Rt(\eb)]}
   \right|_{\eb=-\infty}^{\eb=+\infty} \,\,\, = 0 \, .
 \label{eq-app.terms-vanish}
 \sikie
Further from this convergence, the equations 
(\ref{eq-sec.sf.mf.TRWX-int.rel}) denote 
$I_A(\Omega,\omega) = -J_A(\Omega,\omega)$ whenever the mode 
functions $\psi_{BH,\Omega}^{(\pm)}$ are considered under the 
integration with respect to $\Omega$.


\end{document}